\pdfoutput=1
\documentclass[12pt,a4paper]{article}
\usepackage{ifthen} % for conditional statements
\newboolean{pdflatex}
\setboolean{pdflatex}{true} % False for eps figures 

\newboolean{articletitles}
\setboolean{articletitles}{true} % False removes titles in references

\newboolean{uprightparticles}
\setboolean{uprightparticles}{false} %True for upright particle symbols

\def\paperauthors{LHCb collaboration} % Leave as is for PAPER, CONF and FIGURE
\def\paperasciititle{Measurements of the branching ratio of phi->ee and phi->mumu with charm meson decays} % Set ASCII title here !! MAKE sure it's only ASCII characters !! 
\def\papertitle{Measurements of the branching fraction ratio $\mathcal{B}(\phi\to\mu^{+}\mu^{-})$/$\mathcal{B}(\phi\to e^{+}e^{-})$ with charm meson decays}  % Latex formatted title

\def\paperkeywords{{High Energy Physics}, {LHCb}} % Comma separated list
\def\papercopyright{\the\year\ CERN for the benefit of the LHCb collaboration} % new since 9/Apr/2018
\def\paperlicence{CC BY 4.0 licence}
\def\paperlicenceurl{https://creativecommons.org/licenses/by/4.0/}

\usepackage[top=1in, bottom=1.25in, left=1in, right=1in]{geometry}

\usepackage{microtype}
\usepackage{lineno}  % for line numbering during review
\usepackage{xspace} % To avoid problems with missing or double spaces after
\usepackage{caption} %these three command get the figure and table captions automatically small

\usepackage{graphicx}  % to include figures (can also use other packages)
\usepackage{color}
\usepackage{colortbl}
\graphicspath{{./figs/}} % Make Latex search fig subdir for figures
\usepackage{amsmath} % Adds a large collection of math symbols
\usepackage{amssymb}
\usepackage{amsfonts}
\usepackage{upgreek} % Adds in support for greek letters in roman typeset

\newcommand*\patchAmsMathEnvironmentForLineno[1]{%
\expandafter\let\csname old#1\expandafter\endcsname\csname #1\endcsname
\expandafter\let\csname oldend#1\expandafter\endcsname\csname
end#1\endcsname
 \renewenvironment{#1}%
   {\linenomath\csname old#1\endcsname}%
   {\csname oldend#1\endcsname\endlinenomath}%
}
\newcommand*\patchBothAmsMathEnvironmentsForLineno[1]{%
  \patchAmsMathEnvironmentForLineno{#1}%
  \patchAmsMathEnvironmentForLineno{#1*}%
}
\AtBeginDocument{%
\patchBothAmsMathEnvironmentsForLineno{equation}%
\patchBothAmsMathEnvironmentsForLineno{align}%
\patchBothAmsMathEnvironmentsForLineno{flalign}%
\patchBothAmsMathEnvironmentsForLineno{alignat}%
\patchBothAmsMathEnvironmentsForLineno{gather}%
\patchBothAmsMathEnvironmentsForLineno{multline}%
\patchBothAmsMathEnvironmentsForLineno{eqnarray}%
}

\usepackage{hyperxmp}

\usepackage[pdftex,
            pdfauthor={\paperauthors},
            pdftitle={\paperasciititle},
            pdfkeywords={\paperkeywords},
            pdfcopyright={Copyright (C) \papercopyright},
            pdflicenseurl={\paperlicenceurl}]{hyperref}
\usepackage[colorinlistoftodos,textsize=scriptsize]{todonotes}

\usepackage[bottom,flushmargin,hang,multiple]{footmisc}

\usepackage[all]{hypcap} % Internal hyperlinks to floats.

\usepackage{xspace} 
\usepackage{upgreek}

\def\lhcb   {\mbox{LHCb}\xspace}

\def\MagUp {\mbox{\em Mag\kern -0.05em Up}\xspace}

\ifthenelse{\boolean{uprightparticles}}%
{

 \def\Pmu         {\ensuremath{\upmu}\xspace}

 \def\Ppi         {\ensuremath{\uppi}\xspace}

 \def\Ppsi        {\ensuremath{\uppsi}\xspace}

 \def\PDelta      {\ensuremath{\Delta}\xspace}                 
 \def\PXi         {\ensuremath{\Xi}\xspace}                 
 \def\PLambda     {\ensuremath{\Lambda}\xspace}                 
 \def\PSigma      {\ensuremath{\Sigma}\xspace}                 
 \def\POmega      {\ensuremath{\Omega}\xspace}                 
 \def\PUpsilon    {\ensuremath{\Upsilon}\xspace}
 \let\oldPi\Pi
 \def\PPi         {\ensuremath{\oldPi}\xspace}

 \def\PB      {\ensuremath{\mathrm{B}}\xspace}                 
                  
 \def\PD      {\ensuremath{\mathrm{D}}\xspace}

 \def\PJ      {\ensuremath{\mathrm{J}}\xspace}                 
 \def\PK      {\ensuremath{\mathrm{K}}\xspace}

 \def\Pb      {\ensuremath{\mathrm{b}}\xspace}                 
 \def\Pc      {\ensuremath{\mathrm{c}}\xspace}                 
                  
 \def\Pe      {\ensuremath{\mathrm{e}}\xspace}

 \def\Pi      {\ensuremath{\mathrm{i}}\xspace}

 \def\Ps      {\ensuremath{\mathrm{s}}\xspace}

 \def\thebaroffset{0.0em}
}
{

 \def\Pmu         {\ensuremath{\mu}\xspace}

 \def\Ppi         {\ensuremath{\pi}\xspace}

 \def\Ppsi        {\ensuremath{\psi}\xspace}                 
                  
 \mathchardef\PDelta="7101
 \mathchardef\PXi="7104
 \mathchardef\PLambda="7103
 \mathchardef\PSigma="7106
 \mathchardef\POmega="710A
 \mathchardef\PUpsilon="7107
 \mathchardef\PPi="7105
                  
 \def\PB      {\ensuremath{B}\xspace}                 
                  
 \def\PD      {\ensuremath{D}\xspace}

 \def\PJ      {\ensuremath{J}\xspace}                 
 \def\PK      {\ensuremath{K}\xspace}

 \def\Pb      {\ensuremath{b}\xspace}                 
 \def\Pc      {\ensuremath{c}\xspace}                 
                  
 \def\Pe      {\ensuremath{e}\xspace}

 \def\Pi      {\ensuremath{i}\xspace}

 \def\Ps      {\ensuremath{s}\xspace}

 \def\thebaroffset{0.18em}
}
\newcommand{\offsetoverline}[2][\thebaroffset]{\kern #1\overline{\kern -#1 #2}}%

\makeatletter
\ifcase \@ptsize \relax% 10pt
  \newcommand{\miniscule}{\@setfontsize\miniscule{4}{5}}% \tiny: 5/6
\or% 11pt
  \newcommand{\miniscule}{\@setfontsize\miniscule{5}{6}}% \tiny: 6/7
\or% 12pt
  \newcommand{\miniscule}{\@setfontsize\miniscule{5}{6}}% \tiny: 6/7
\fi
\makeatother

\DeclareRobustCommand{\optbar}[1]{\shortstack{{\miniscule (\rule[.5ex]{1.25em}{.18mm})}
  \\ [-.7ex] $#1$}}

\def\epem       {{\ensuremath{\Pe^+\Pe^-}}\xspace}
\def\mumu       {{\ensuremath{\Pmu^+\Pmu^-}}\xspace}

\def\ellell     {\ensuremath{\ell^+ \ell^-}\xspace}

\def\squark    {{\ensuremath{\Ps}}\xspace}

\def\cquark    {{\ensuremath{\Pc}}\xspace}

\def\bquark    {{\ensuremath{\Pb}}\xspace}

\def\pion   {{\ensuremath{\Ppi}}\xspace}

\def\pip    {{\ensuremath{\pion^+}}\xspace}
\def\pim    {{\ensuremath{\pion^-}}\xspace}

\def\kaon    {{\ensuremath{\PK}}\xspace}
\def\KorKbar {\kern \thebaroffset\optbar{\kern -\thebaroffset \PK}{}\xspace}

\def\Kp      {{\ensuremath{\kaon^+}}\xspace}

\def\D       {{\ensuremath{\PD}}\xspace}

\def\DorDbar {\kern \thebaroffset\optbar{\kern -\thebaroffset \PD}\xspace}
\def\Dz      {{\ensuremath{\D^0}}\xspace}

\def\Dp      {{\ensuremath{\D^+}}\xspace}
\def\Dm      {{\ensuremath{\D^-}}\xspace}

\def\DpDm    {\ensuremath{\Dp {\kern -0.16em \Dm}}\xspace}

\def\Dstarp  {{\ensuremath{\D^{*+}}}\xspace}

\def\Ds      {{\ensuremath{\D^+_\squark}}\xspace}

\def\B       {{\ensuremath{\PB}}\xspace}

\def\BorBbar {\kern \thebaroffset\optbar{\kern -\thebaroffset \PB}\xspace}

\def\Bd      {{\ensuremath{\B^0}}\xspace}

\def\BdorBdbar {\kern \thebaroffset\optbar{\kern -\thebaroffset \Bd}\xspace}
\def\Bu      {{\ensuremath{\B^+}}\xspace}

\def\Bp      {{\ensuremath{\Bu}}\xspace}

\def\Bs      {{\ensuremath{\B^0_\squark}}\xspace}

\def\BsorBsbar {\kern \thebaroffset\optbar{\kern -\thebaroffset \Bs}\xspace}

\def\jpsi     {{\ensuremath{{\PJ\mskip -3mu/\mskip -2mu\Ppsi}}}\xspace}
\def\psitwos  {{\ensuremath{\Ppsi{(2S)}}}\xspace}

\def\Y#1S{\ensuremath{\PUpsilon{(#1S)}}\xspace}

\def\LorLbar     {\kern \thebaroffset\optbar{\kern -\thebaroffset \PLambda}\xspace}

\newcommand{\decay}[2]{\ensuremath{#1\!\to #2}\xspace} 

\def\to                 {\ensuremath{\rightarrow}\xspace}

\def\qsq       {{\ensuremath{q^2}}\xspace}

\def\bsll     {\decay{\bquark}{\squark \ell^+ \ell^-}}

\def\AT#1     {\ensuremath{A_{\mathrm{T}}^{#1}}\xspace}           % 2

\def\C#1      {\ensuremath{\mathcal{C}_{#1}}\xspace}                       % 9
\def\Cp#1     {\ensuremath{\mathcal{C}_{#1}^{'}}\xspace}                    % 7
\def\Ceff#1   {\ensuremath{\mathcal{C}_{#1}^{\mathrm{(eff)}}}\xspace}        % 9  
\def\Cpeff#1  {\ensuremath{\mathcal{C}_{#1}^{'\mathrm{(eff)}}}\xspace}       % 7
\def\Ope#1    {\ensuremath{\mathcal{O}_{#1}}\xspace}                       % 2
\def\Opep#1   {\ensuremath{\mathcal{O}_{#1}^{'}}\xspace}                    % 7

\newcommand{\nospaceunit}[1]{\ensuremath{\text{#1}}}       
\newcommand{\aunit}[1]{\ensuremath{\text{\,#1}}}       
\newcommand{\tev}{\aunit{Te\kern -0.1em V}\xspace}
\newcommand{\gev}{\aunit{Ge\kern -0.1em V}\xspace}
\newcommand{\mev}{\aunit{Me\kern -0.1em V}\xspace}
\newcommand{\kev}{\aunit{ke\kern -0.1em V}\xspace}
\newcommand{\ev}{\aunit{e\kern -0.1em V}\xspace}
 
\newcommand{\mevc}{\ensuremath{\aunit{Me\kern -0.1em V\!/}c}\xspace}
\newcommand{\gevc}{\ensuremath{\aunit{Ge\kern -0.1em V\!/}c}\xspace}
\newcommand{\mevcc}{\ensuremath{\aunit{Me\kern -0.1em V\!/}c^2}\xspace}
\newcommand{\gevcc}{\ensuremath{\aunit{Ge\kern -0.1em V\!/}c^2}\xspace}
\def\mum  {\ensuremath{\,\upmu\nospaceunit{m}}\xspace}

\def\fb   {\ensuremath{\aunit{fb}}\xspace}
\def\invfb   {\ensuremath{\fb^{-1}}\xspace}

\newcommand{\chisq}{\ensuremath{\chi^2}\xspace}

\def\gsim{{~\raise.15em\hbox{$>$}\kern-.85em
          \lower.35em\hbox{$\sim$}~}\xspace}
\def\lsim{{~\raise.15em\hbox{$<$}\kern-.85em
          \lower.35em\hbox{$\sim$}~}\xspace}

\def\sPlot{\mbox{\em sPlot}\xspace}

\def\pt         {\ensuremath{p_{\mathrm{T}}}\xspace}

\def\ptot       {\ensuremath{p}\xspace}

\def\evtgen     {\mbox{\textsc{EvtGen}}\xspace}

\def\geant      {\mbox{\textsc{Geant4}}\xspace}

\def\photos     {\mbox{\textsc{Photos}}\xspace}

\def\pythia     {\mbox{\textsc{Pythia}}\xspace}

\def\tell1  {TELL1\xspace}
\def\ukl1   {UKL1\xspace}

\newcommand{\lhcborcid}[1]{\href{https://orcid.org/#1}{\hspace*{0.1em}\raisebox{-0.45ex}{\includegraphics[width=1em]{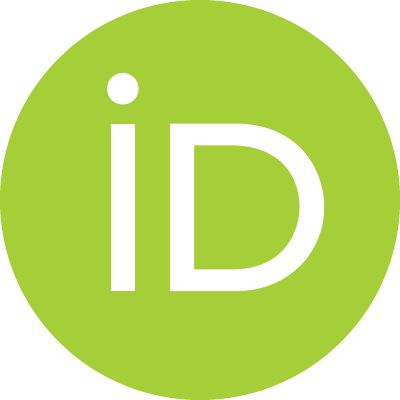}}}}

\usepackage{cite} % Allows for ranges in citations
\usepackage{mciteplus}
\usepackage{booktabs} 

\def\phimumu{{\ensuremath{\phi\to \mumu}}\xspace}
\def\phiee{{\ensuremath{\phi\to \epem}}\xspace}

\def\PiPhiee{{\ensuremath{\pip \phi(\to e^{+}e^{-})}}\xspace}
\def\PiPhimumu{{\ensuremath{\pip \phi(\to \mu^{+}\mu^{-})}}\xspace}

\def\DsPiPhi{{\ensuremath{\Ds \to \pip \phi}}\xspace}
\def\DpPiPhi{{\ensuremath{\Dp \to \pip \phi}}\xspace}
\def\DPiPhi{{\ensuremath{D^{+}_{(s)} \to \pip \phi}}\xspace}

\def\DPiPiPi{{\ensuremath{D^{+}_{(s)}\to \pip\pim\pip\xspace}}}

\def\DsPiPhiee{{\ensuremath{{\Ds \to \pip \phi(\to \epem)}}}\xspace}
\def\DpPiPhiee{{\ensuremath{{\Dp \to \pip \phi(\to \epem)}}}\xspace}
\def\DPiPhiee{{\ensuremath{{D^{+}_{(s)} \to \pip \phi(\to \epem)}}}\xspace}

\def\DsPiPhimumu{{\ensuremath{{\Ds \to \pip \phi (\to \mumu)}}}\xspace}
\def\DpPiPhimumu{{\ensuremath{{\Dp \to \pip \phi (\to \mumu)}}}\xspace}
\def\DPiPhimumu{{\ensuremath{{D^{+}_{(s)} \to \pip \phi (\to \mumu)}}}\xspace}

\def\BKJPsi{{\ensuremath{\Bp\to K^{+} \jpsi }}\xspace}
\def\BKJPsill{{\ensuremath{{\Bp\to K^{+} \jpsi(\to \ell^{+}\ell^{-})}}}\xspace}
\def\BKJPsiee{{\ensuremath{{\Bp\to K^{+}\jpsi(\to \epem)}}}\xspace}
\def\BKJPsimumu{{\ensuremath{{\Bp\to K^{+}\jpsi(\to \mumu)}}}\xspace}

\def\BHJPsiee{{\ensuremath{\Bp\to H_{s}  \jpsi ( \to \epem) }}\xspace}
\def\BHJPsimumu{{\ensuremath{\Bp\to H_{s} \jpsi ( \to \mumu) }}\xspace}

\def\Rbothphipi{{\ensuremath{R^{(d,s)}_{\phi \pi}}}\xspace}
\def\Rsphipi{{\ensuremath{R^{s}_{\phi \pi}}}\xspace}
\def\Rdphipi{{\ensuremath{R^{d}_{\phi \pi}}}\xspace}
\def\Rphipi{{\ensuremath{R_{\phi \pi}}}\xspace}

\def\rjpsi{{\ensuremath{r_{\jpsi}}}\xspace}
\def\rpsitwos{{\ensuremath{R_{\psitwos}}}\xspace}

\def\bsll{{\ensuremath{b\to s \ell^{+}\ell^{-}}}\xspace}

\def\BdorBu  {{\ensuremath{\B^{(0,+)}}}\xspace}
\def\mKllconst  {\ensuremath{m_{\jpsi}{(\Kp\ellell)}}\xspace}

\def\Pill  {\ensuremath{\pip\ellell}\xspace}
\def\mPillconst  {\ensuremath{m_{\phi}{(\Pill)}}\xspace}

\def\BuBdKpijpsi {\decay{\BdorBu}{\Kp\pi^{(-,0)} \jpsi(\to\epem)}\xspace}

\def\RdphipiResult {\ensuremath{\Rdphipi &= 1.026 \pm 0.020 \,({\rm stat}) \, \pm 0.056 \,({\rm syst})}\xspace}
\def\RsphipiResult {\ensuremath{\Rsphipi &= 1.017 \pm 0.013 \,({\rm stat}) \, \pm 0.051 \,({\rm syst})}\xspace}
\def\RphipiResult {\ensuremath{\Rphipi = 1.022 \pm 0.012 \,({\rm stat}) \, \pm 0.048 \,({\rm syst})}\xspace}
\usepackage{longtable} % only for template; not usually to be used in PAPERs
\usepackage{multirow}
\begin{document}

%%%%%%%%%%%%%%%%%%%%%%%%%
%%%%% Title     %%%%%%%%%
%%%%%%%%%%%%%%%%%%%%%%%%%
\renewcommand{\thefootnote}{\fnsymbol{footnote}}
\setcounter{footnote}{1}

% %%%%%%% CHOOSE TITLE PAGE--------
%\onecolumn
%\input{title-LHCb-INT}
%\input{title-LHCb-ANA}
%\input{title-LHCb-CONF}
%\input{title-LHCb-FIGURE}
% ===============================================================================
% Purpose: LHCb-PAPER journal paper title page template
% Author: 
% Created on: 2010-09-25
% ===============================================================================

%%%%%%%%%%%%%%%%%%%%%%%%%
%%%%%  TITLE PAGE  %%%%%%
%%%%%%%%%%%%%%%%%%%%%%%%%
\begin{titlepage}
\pagenumbering{roman}

% Header ---------------------------------------------------
\vspace*{-1.5cm}
\centerline{\large EUROPEAN ORGANIZATION FOR NUCLEAR RESEARCH (CERN)}
\vspace*{1.5cm}
\noindent
\begin{tabular*}{\linewidth}{lc@{\extracolsep{\fill}}r@{\extracolsep{0pt}}}
\ifthenelse{\boolean{pdflatex}}% Logo format choice
{\vspace*{-1.5cm}\mbox{\!\!\!\includegraphics[width=.14\textwidth]{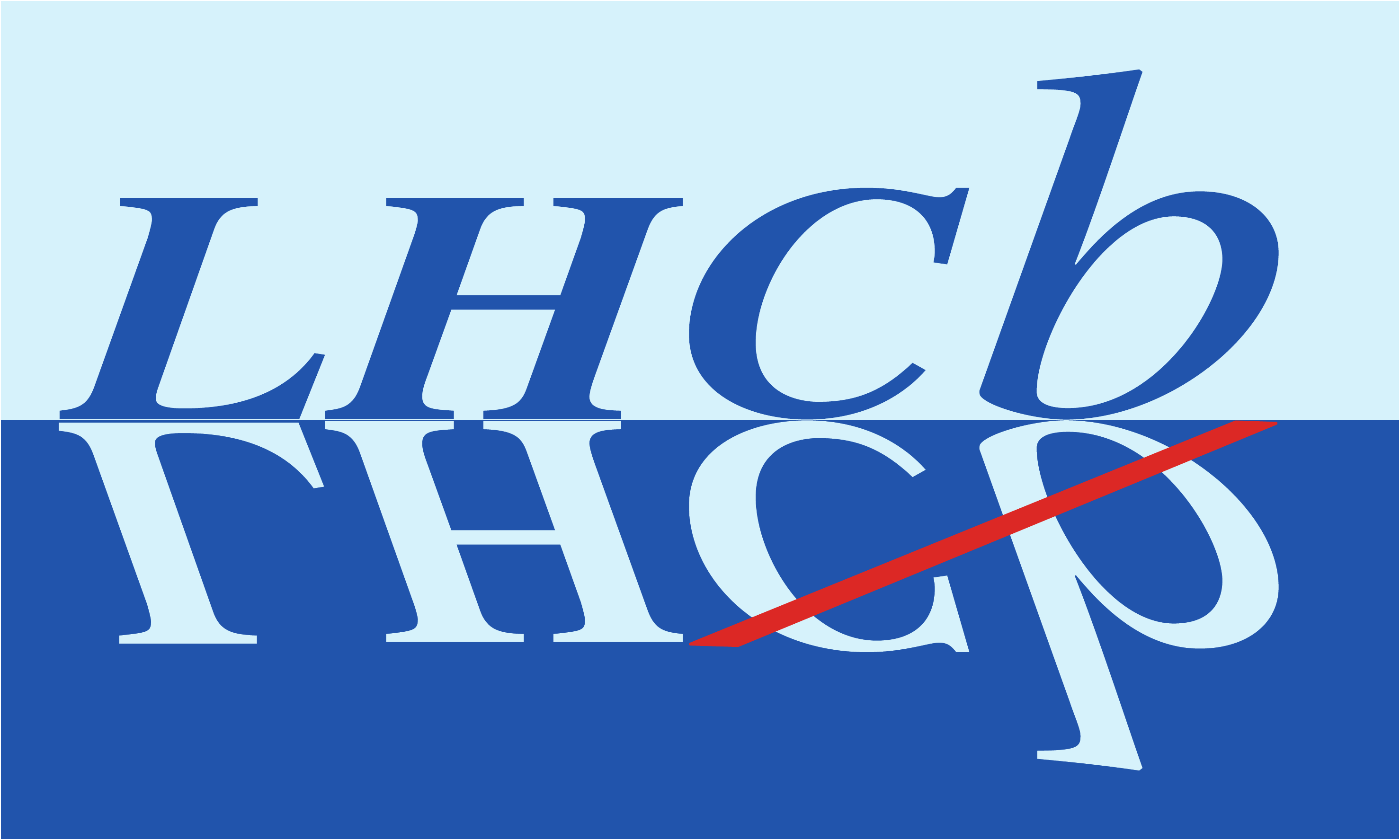}} & &}%
{\vspace*{-1.2cm}\mbox{\!\!\!\includegraphics[width=.12\textwidth]{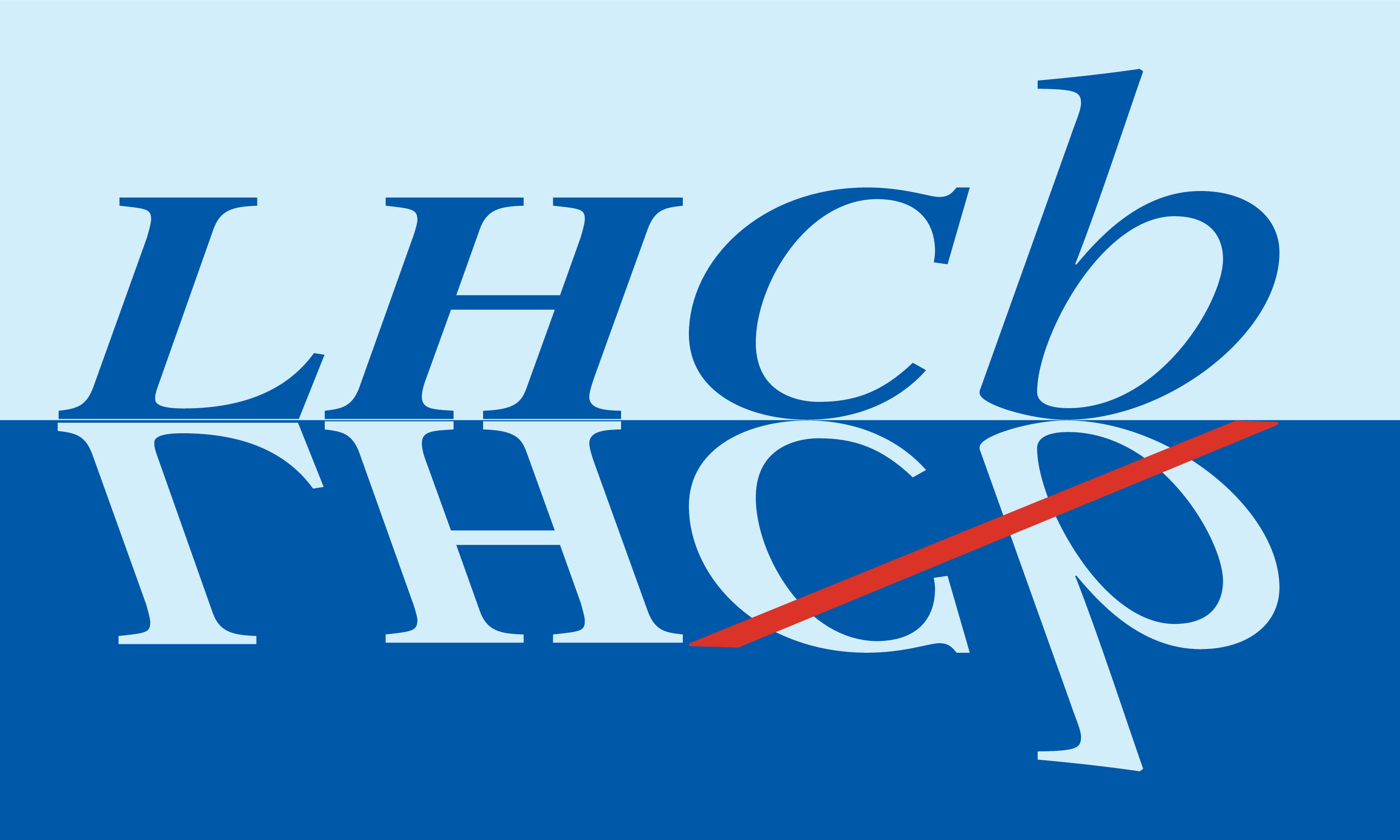}} & &}%
\\
 & & CERN-EP-2024-001 \\  % ID 
 & & LHCb-PAPER-2023-038 \\  % ID 
 %& & \today \\ % Date
 & & August 9, 2024 \\ % Date
 & & \\
% not in paper \hline
\end{tabular*}

\vspace*{3.0cm}

% Title --------------------------------------------------
{\normalfont\bfseries\boldmath\huge
\begin{center}
% DO NOT EDIT HERE. Instead edit macro in main.tex to keep metadata correct
  \papertitle
\end{center}
}

\vspace*{2.0cm}

% Authors -------------------------------------------------
\begin{center}
%In the footnote, replace 'paper' by 'Letter' in case of submission to PRL or PLB 
% Edit macro in main.tex to keep metadata correct
\paperauthors\footnote{Authors are listed at the end of this paper.}
\end{center}

\vspace{\fill}

% Abstract -----------------------------------------------
\begin{abstract}
  \noindent
  Measurements of the branching fraction ratio $\mathcal{B}$(\phimumu)/$\mathcal{B}$(\phiee) with \DsPiPhi and \DpPiPhi decays, denoted \Rsphipi and \Rdphipi, are presented. The analysis is performed using a dataset corresponding to an integrated luminosity of  5.4\invfb of $pp$ collision data collected with the LHCb experiment. The branching fractions are normalised with respect to the \BKJPsiee and \BKJPsimumu decay modes. The combination of the results yields
  \begin{displaymath}
  \RphipiResult .
  \end{displaymath}
The result is compatible with previous measurements of the $\phi \to \ell^{+}\ell^{-}$ branching fractions and predictions based on the Standard Model.
  
\end{abstract}

\vspace*{1.0cm}

\begin{center}
  Published in JHEP 05 (2024) 293
\end{center}

\vspace{\fill}

{\footnotesize 
% Edit macro in main.tex to keep metadata correct
\centerline{\copyright~\papercopyright. \href{\paperlicenceurl}{\paperlicence}.}}
\vspace*{2mm}

\end{titlepage}

%%%%%%%%%%%%%%%%%%%%%%%%%%%%%%%%
%%%%%  EOD OF TITLE PAGE  %%%%%%
%%%%%%%%%%%%%%%%%%%%%%%%%%%%%%%%

%  empty page follows the title page ----
\newpage
\setcounter{page}{2}
\mbox{~}
%\newpage
%
%% Author List ----------------------------
%%  You need to get a new author list!
%\input{LHCb_authorlist.tex}
%
%The author list for journal publications is provided by the Membership Committee shortly after 'approval to go to paper' has been given.
%%It will be made available on the page
%%\verb!http://www.physik.uzh.ch/~strauman/forMemCo/LHCb-PAPER-XXXX-XXX/! .
%It will be sent to you by email shortly after a paper number has beens assigned.
%The author list should be included already at first circulation, 
%to allow new members of the collaboration to verify whether they have been included correctly.
%Occasionally a misspelled name is corrected or associated institutions become full members.
%In that case, a new author list will be sent to you.
%In case line numbering doesn't work well after including the authorlist, try moving the \verb!\bigskip! after the last author to a separate line.
%
%
%The authorship for Conference Reports should be ``The LHCb
%  collaboration'', with a footnote giving the name(s) of the contact
%  author(s), but without the full list of collaboration names.

%\twocolumn
% %%%%%%%%%%%%% ---------

\renewcommand{\thefootnote}{\arabic{footnote}}
\setcounter{footnote}{0}

%%%%%%%%%%%%%%%%%%%%%%%%%%%%%%%%
%%%%%  Table of Content   %%%%%%
%%%%%%%%%%%%%%%%%%%%%%%%%%%%%%%%
%%%% Uncomment if desired
%\tableofcontents
\cleardoublepage

%%%%%%%%%%%%%%%%%%%%%%%%%
%%%%% Main text %%%%%%%%%
%%%%%%%%%%%%%%%%%%%%%%%%%

\pagestyle{plain} % restore page numbers for the main text
\setcounter{page}{1}
\pagenumbering{arabic}

%% Uncomment during review phase. 
%% Comment before a final submission.
%\linenumbers

%% This is the main body
%% It is useful to have a single file so comemnts are not missed in overleaf.
\section{Introduction}
\label{sec:intro}

A particularly sensitive way of searching for physics beyond the Standard Model (SM) is to make high-precision experimental tests of accidental symmetries that arise in the SM. One such accidental symmetry is lepton flavour universality (LFU), whereby the strength of the electroweak couplings for charged leptons is independent of their flavour. The violation of LFU is a common feature in extensions of the SM, and branching fraction ratios of hadron decays  involving different lepton species constitute a sensitive probe for physics beyond the SM, since they are largely free from hadronic uncertainties.

There has been much activity in testing LFU in $b$-hadron decays over the last decade at LHCb. In \bsll decays, where $\ell$ represents either an electron or a muon, the most recent LFU measurements~\cite{LHCb-PAPER-2022-046,LHCb-PAPER-2022-045} are consistent with the SM with a precision of 5\% that is statistically limited. Several cross-checks are performed in the \bsll LFU analyses to reinforce confidence in the results. The most stringent of these checks is the ratio ${\rjpsi=\mathcal{B}(\BHJPsimumu)/\mathcal{B}(\BHJPsiee)}$, where $H_{s}$ represents a strange meson. The measurement of the ratio ${\rjpsi}$ requires significantly more control over the efficiencies than the LFU measurements themselves and was found to be compatible with the SM prediction in all measurements~\cite{LHCB-PAPER-2014-024,LHCB-PAPER-2017-013,LHCb-PAPER-2019-009,LHCb-PAPER-2021-004,LHCb-PAPER-2021-038}. The variation of the efficiency as a function of several kinematic variables, such as the maximum transverse momentum of the two leptons, is also performed and no significant systematic effects are seen. Finally, the double ratio for the \psitwos and \jpsi resonances, \rpsitwos, is performed and is also found to be compatible with unity for all measurements~\cite{LHCB-PAPER-2014-024,LHCB-PAPER-2017-013,LHCb-PAPER-2019-009,LHCb-PAPER-2021-004,LHCb-PAPER-2021-038}. The robustness of the efficiency for the existing LFU measurements at the current level of precision is therefore clear. However, as the statistical precision improves in the future, analysing a control channel in the \qsq region complementary to those performed in the existing analyses will be important to keep systematic uncertainties subdominant. 

The branching fraction ratios 
\begin{equation}
    \Rbothphipi \equiv \frac{\mathcal{B}(\DPiPhimumu)}{\mathcal{B}(\DPiPhiee)},
\end{equation}
\noindent where $\phi$ refers to the $\phi(1020)$ meson, are ideal quantities in this respect~\cite{Altmannshofer:2017bsz}. As the decay is dominated by photon exchange, LFU is expected to hold in this process.  The invariant mass of the $\phi$ meson means that this decay occurs at a \qsq value below many of the existing LFU measurements. The decays \DPiPhi have a large branching fraction~\cite{PDG2022} and a fully charged final state, meaning that high precision tests can be performed.\footnote{The inclusion of the charge-conjugate mode is implied.} Furthermore, the current precision of $\mathcal{B}$(\phimumu)/$\mathcal{B}$(\phiee) is 6\%, obtained by comparing the individual branching fraction measurements from Refs.~\cite{Akhmetshin:1995vz,Akhmetshin:2010dt,KLOE:2004uzx}. This test is entirely limited by the branching fraction of $\mathcal{B}$(\phimumu)~\cite{KLOE:2004uzx}, which provides the opportunity to improve the precision of that branching fraction.

This paper presents a measurement of $\mathcal{B}$(\phimumu)/$\mathcal{B}$(\phiee) based on a dataset corresponding to an integrated luminosity of 5.4\invfb of $pp$ collisions collected with the LHCb detector between 2016 and 2018. Similarly to the $\bsll$ LFU measurements performed by the LHCb collaboration, the signals are measured with respect to the \BKJPsill decay modes.
\noindent The double ratio strategy, which consists in comparing the branching fraction ratios of signal and normalisation modes with the same topology, ensures that the knowledge of absolute efficiencies is unnecessary. This normalisation assumes that LFU holds in \jpsi decays, which has been verified well beyond the precision of this measurement~\cite{PDG2022}.

The paper is organised as follows: firstly, a brief overview of the LHCb experiment is given in Sect.~\ref{sec:detector}, along with a description of the dataset. Secondly, the signal selection is given in Sect.~\ref{sec:selection} and the remaining backgrounds and signal-yield determination are discussed in Sect.~\ref{sec:fit}. The determination of the reconstruction and selection efficiency of the signal and normalisation modes is presented in Sect.~\ref{sec:efficiency}. The values of these efficiencies were kept blind until the analysis was finalised to avoid experimenter bias. Sources of systematic uncertainties and their impact are discussed in Sect.~\ref{sec:systematics}. Finally, the results and conclusion are presented in Sect.~\ref{sec:results}.

\section{The LHCb detector and simulation}
\label{sec:detector}

The \lhcb detector~\cite{LHCb-DP-2008-001,LHCb-DP-2014-002} is a single-arm forward
spectrometer covering the \mbox{pseudorapidity} range $2<\eta <5$,
designed for the study of particles containing \bquark or \cquark
quarks. The detector includes a high-precision tracking system
consisting of a silicon-strip vertex detector surrounding the $pp$
interaction region~\cite{LHCb-DP-2014-001}, a large-area silicon-strip detector located
upstream of a dipole magnet with a bending power of about
$4{\mathrm{\,Tm}}$, and three stations of silicon-strip detectors and straw
drift tubes~\cite{LHCb-DP-2017-001}
placed downstream of the magnet.
The tracking system provides a measurement of the momentum, \ptot, of charged particles with
a relative uncertainty that varies from 0.5\% at low momentum to 1.0\% at 200\gevc.
The minimum distance of a track to a primary $pp$ collision vertex (PV), the impact parameter (IP), 
is measured with a resolution of $(15+29/\pt)\mum$,
where \pt is the component of the momentum transverse to the beam, in\,\gevc.
Different types of charged hadrons are distinguished using information
from two ring-imaging Cherenkov detectors~\cite{LHCb-DP-2012-003}. 
Photons, electrons and hadrons are identified by a calorimeter system consisting of
scintillating-pad (SPD) and preshower detectors, an electromagnetic
% calorimeter
and a hadronic calorimeter. Muons are identified by a
system composed of alternating layers of iron and multiwire
proportional chambers~\cite{LHCb-DP-2012-002}.

Simulation is required to model the effects of the detector acceptance and to account for the
  imposed selection requirements.
  In the simulation, $pp$ collisions are generated using
  \pythia~\cite{Sjostrand:2007gs,*Sjostrand:2006za} 
  with a specific \lhcb configuration~\cite{LHCb-PROC-2010-056}.
  Decays of unstable particles
  are described by \evtgen~\cite{Lange:2001uf}, in which final-state
  radiation is generated using \photos~\cite{davidson2015photos}. The
  interaction of the generated particles with the detector, and its response,
  are implemented using the \geant
  toolkit~\cite{Allison:2006ve, *Agostinelli:2002hh}, as described in
  Ref.~\cite{LHCb-PROC-2011-006}.

\section{Candidate selection}
\label{sec:selection}

The online event selection is performed by a trigger~\cite{LHCb-DP-2012-004}, 
which consists of a hardware stage, based on information from the calorimeter and muon systems, followed by a software stage, which applies a full event
reconstruction. At the hardware trigger stage, candidates with muons are required to have at least one high-\pt muon track. Decays involving electrons are required to either include a high-\pt electron, a high-\pt hadron or be triggered independently of the final state. The data selected by these three trigger methods in the electron channel are treated as separate categories in the analysis. 
  
The software trigger is split into two stages, with the first requiring one or two tracks with high IP and \pt. In between the two software stages, an alignment and calibration of the detector is performed in near real-time and the results are used in the second stage of the software trigger~\cite{LHCb-PROC-2015-011}.
      The same alignment and calibration information is propagated 
      to the offline reconstruction, ensuring that the 
      particle identification (PID) is of high quality and consistent between the trigger and 
      offline analysis. The identical performance of the online and offline 
      reconstruction enables physics analyses to be 
      performed directly using candidates reconstructed in the trigger 
      \cite{LHCb-DP-2012-004,LHCb-DP-2016-001}, 
      which the present analysis exploits for the signal channels.

In the second software stage, exclusive selection requirements designed to select $D^{+}_{(s)}\to  h^{+}\ell^{+}\ell^{-}$ candidates are applied. Three tracks are required to be inconsistent with originating from any PV, to have $\pt > 300\mevc$ and $p > 2000\mevc$. These tracks are combined to form a good-quality vertex that is significantly displaced with respect to any PV. For each electron track, a search is therefore made in the ECAL for bremsstrahlung energy deposits around the extrapolated track direction before the magnet that are not associated with any other charged tracks. The energy of any such deposit is added to the electron energy that is derived from the measurements made in the tracking system. Bremsstrahlung photons can be added to none, either, or both of the final-state $e^+$ and $e^-$ candidates. The invariant mass of the reconstructed final state particles is required to fall within 200\mevcc of the known \Dp mass~\cite{PDG2022}. The angle between the momentum vector of the charm meson candidate and the line connecting the PV to the decay vertex must be less than 14 mrad. To select the $\phi\to \ell^{+}\ell^{-}$ decay, a dilepton mass window of $990$\,--\,$1050\mevcc$ for muons and $870$\,--\,$1110\mevcc$ for electrons is applied.

The trigger strategy for the normalisation mode \BKJPsi is an inclusive selection that forms two- and three-track combinations with a high-quality vertex that is significantly displaced from any PV. The \BKJPsi candidates are then built offline, requiring that a well-identified kaon is combined with two leptons to form a high-quality vertex. These candidates must be consistent with originating from a PV and their decay vertex must be significantly displaced from that PV. In order to select the \jpsi resonance, the dilepton invariant mass is required to be between $2946$\,--\,$3176\mevcc$ for muons and $2450$\,--\,$3600\mevcc$ for electrons. 

The offline selection encompasses invariant mass vetoes and multivariate analysis techniques. Doubly misidentified background from ${D^{+} \to K^{-}\pi^{+}\pi^{+}}$ decays, where a kaon and a pion are reconstructed as electrons, is rejected by removing candidates that have a $K^{-}\pi^{+}\pi^{+}$ invariant mass within 30\mevcc of the \Dp mass~\cite{PDG2022}. Background sources coming from doubly misidentified $\DPiPiPi$ decays with two pions misidentified as electrons are reduced by the use of particle-identification (PID) criteria. The remaining contamination from this source is included in the fit used to determine the signal yield as described in Sect.~\ref{sec:fit}. 
A kinematic fit that constrains the dilepton invariant mass to the known $\phi$ ($\jpsi$) mass~\cite{PDG2022} is performed to each candidate of the signal (normalisation) mode. In the following, these quantities are referred to as \mPillconst and \mKllconst, respectively. Their distributions are used to extract the signal and normalisation yields via maximum-likelihood fits to the data and to define the sideband regions, summarised in Table~\ref{tab:sidebandranges}. The dominant source of background is combinatorial in nature, whereby the tracks do not originate from a common particle decay. To reduce this background for both signal and normalisation modes, separate Boosted Decision Tree (BDT) classifiers~\cite{Breiman,AdaBoost} are trained. The BDT classifiers use a combination of kinematic and geometric information such as the final state impact parameters and kinematics. The most discriminating variable for both signal and normalisation modes is the angle between the momentum vector of the $D_{(s)}^{+}$ or \Bp and the line connecting the PV with their decay vertices.

\begin{table}[t]
\centering
\caption{Signal (\DPiPhi) and normalisation (\BKJPsill) mode sideband mass ranges. The variables \mPillconst and \mKllconst are used for signal and normalisation decays, respectively.}
\label{tab:sidebandranges}
\begin{tabular}{ccc}
\toprule
Decay mode  & \mPillconst & \mKllconst  \\
            & $[\mevcc]$  & $[\mevcc]$  \\
\midrule
{\begin{tabular}{r@{\;}l}&$e^+e^-$\\&$\mu^+\mu^-$\\\end{tabular}}   & {\begin{tabular}{@{\,}l}\;$\notin$\;[1810,2040]\\ \;$\notin$\;[1840,2000]\end{tabular}} & 
{\begin{tabular}{@{\,}l} \;$>$\;5580\\ \;$>$\;5480\end{tabular}}\\
\bottomrule
\end{tabular}
\end{table}

The requirement on the response of the BDT classifier for the signal mode is optimised using the Punzi figure of merit~\cite{Punzi:2003bu} $N_{s}/(\sigma/2 + \sqrt{N_b})$, where $\sigma=5$, $N_{s}$ is the number of signal candidates as determined from simulation, and $N_b$ is the number of background events extrapolated from the sideband regions for the signal modes. The efficiency for the BDT classifier selection is 59.2\% (70.8\%) on the electron (muon) signal with a 92.8\% (85.4\%) rejection rate for the electron (muon) background. To optimise the BDT classifier response in normalisation mode decays, the metric $N_{s}/\sqrt{N_b}$ is used instead. The use of this metric results in a very pure sample, which simplifies kinematic comparisons between data and simulation. The signal selection efficiency in this case is 99.2\% (92.3\%) on the electron (muon) signal, with a 81.5\% (87.0\%) rejection rate for the electron (muon) background.

\section{Signal yield determination}
\label{sec:fit}

\begin{figure}[ht!]
\centering
\includegraphics[width=0.46\textwidth]{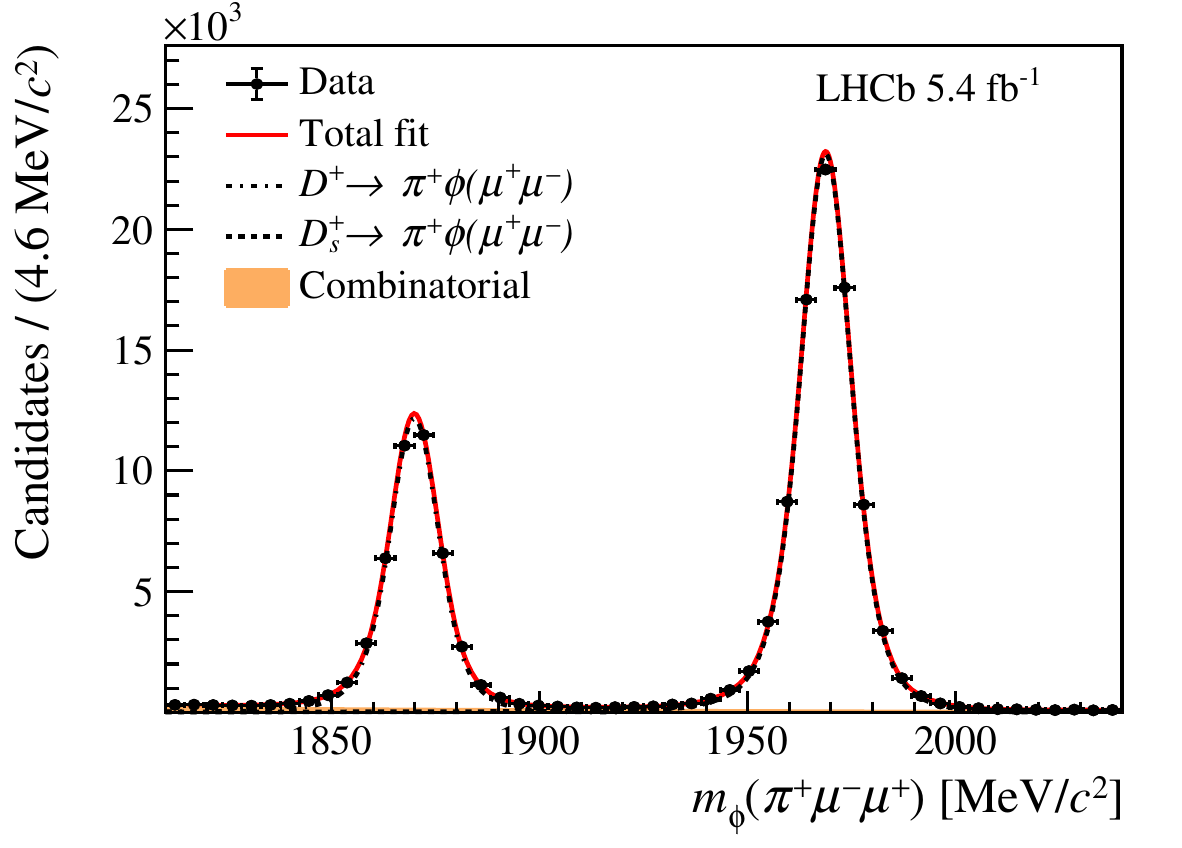}
\includegraphics[width=0.46\textwidth]{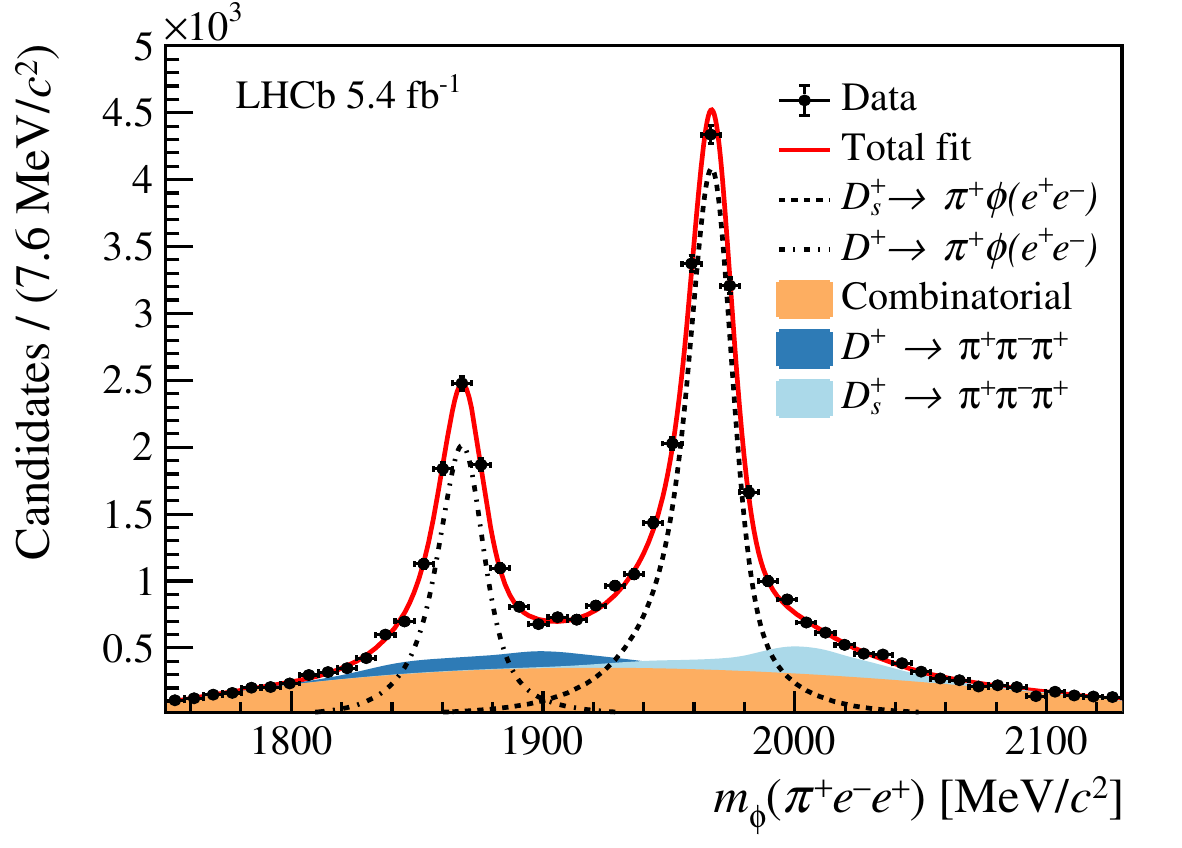}
\caption{Distribution of the invariant mass \mPillconst overlaid with the result of the fit for candidates with (left) muon and (right) electron pairs in the final states. The dotted and dot-dashed lines describe the signal contributions, while the solid areas represent the background components. }
\label{fig:signalfits}
\end{figure}

The signal yield is determined in categories of the different run periods and trigger conditions by means of a maximum-likelihood fit to the \mPillconst distributions of the signal candidates. The $\phi$ mass constraint significantly improves the invariant-mass resolution in the electron channels and leaves the muon channels largely unchanged. The result of the fit is shown in Fig.~\ref{fig:signalfits}.

The signal lineshapes for the electron modes are described by the sum of two Crystal Ball functions~\cite{Skwarnicki:1986xj} with common mean peak position and different width and tail parameters. The lineshapes are determined separately in three bremsstrahlung categories, depending on whether none, one or both electrons are associated with bremsstrahlung photons. The width and tail parameters of these distributions, as well as the fractions of each bremsstrahlung category, are determined from simulation. In order to account for residual differences in the signal shape between data and simulation, an offset in the peak position and a scaling of the resolution are allowed to vary in the fits to the data. The muon decay mode is described with a Gaussian and the sum of two Crystal Ball functions. The shape parameters of these distributions are determined from simulation, with an offset in the peak position and scaling of the resolution allowed.

\begin{figure}[b]
\centering
\includegraphics[width=0.48\textwidth]{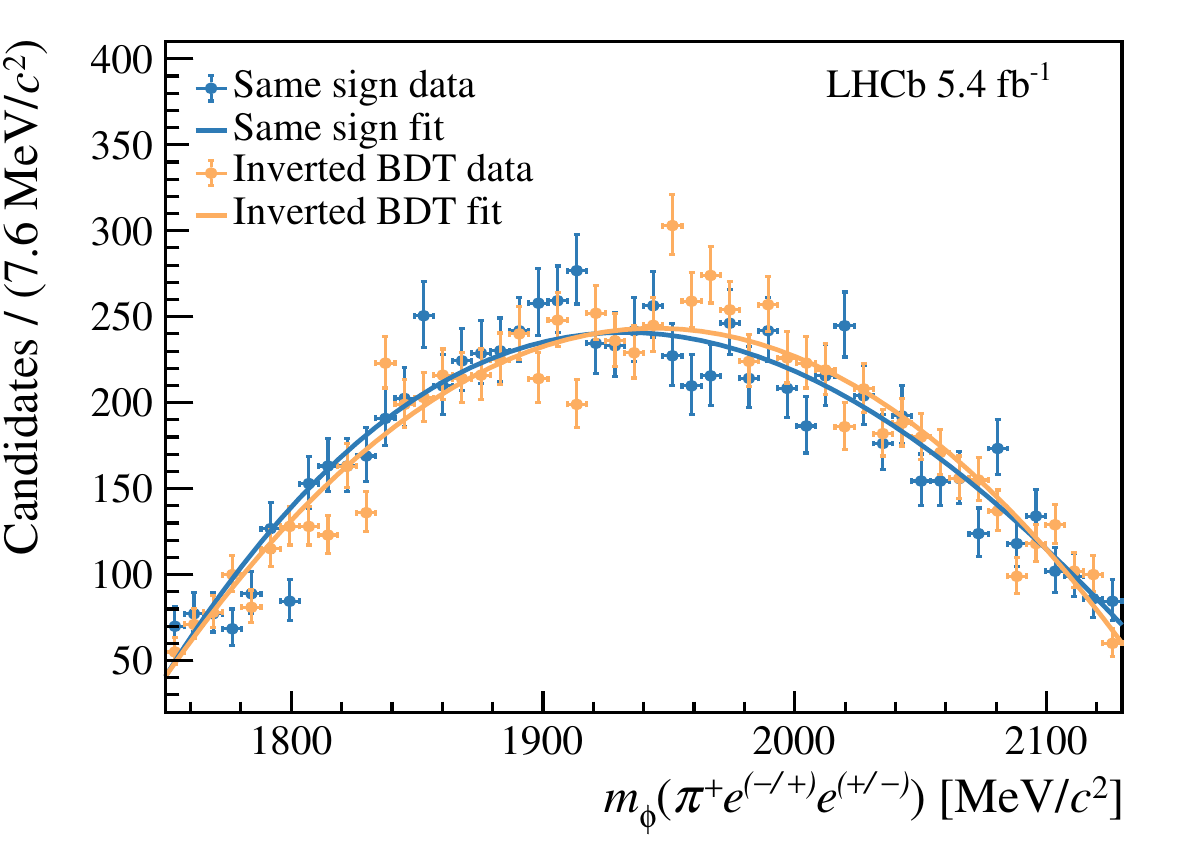}
\includegraphics[width=0.48\textwidth]{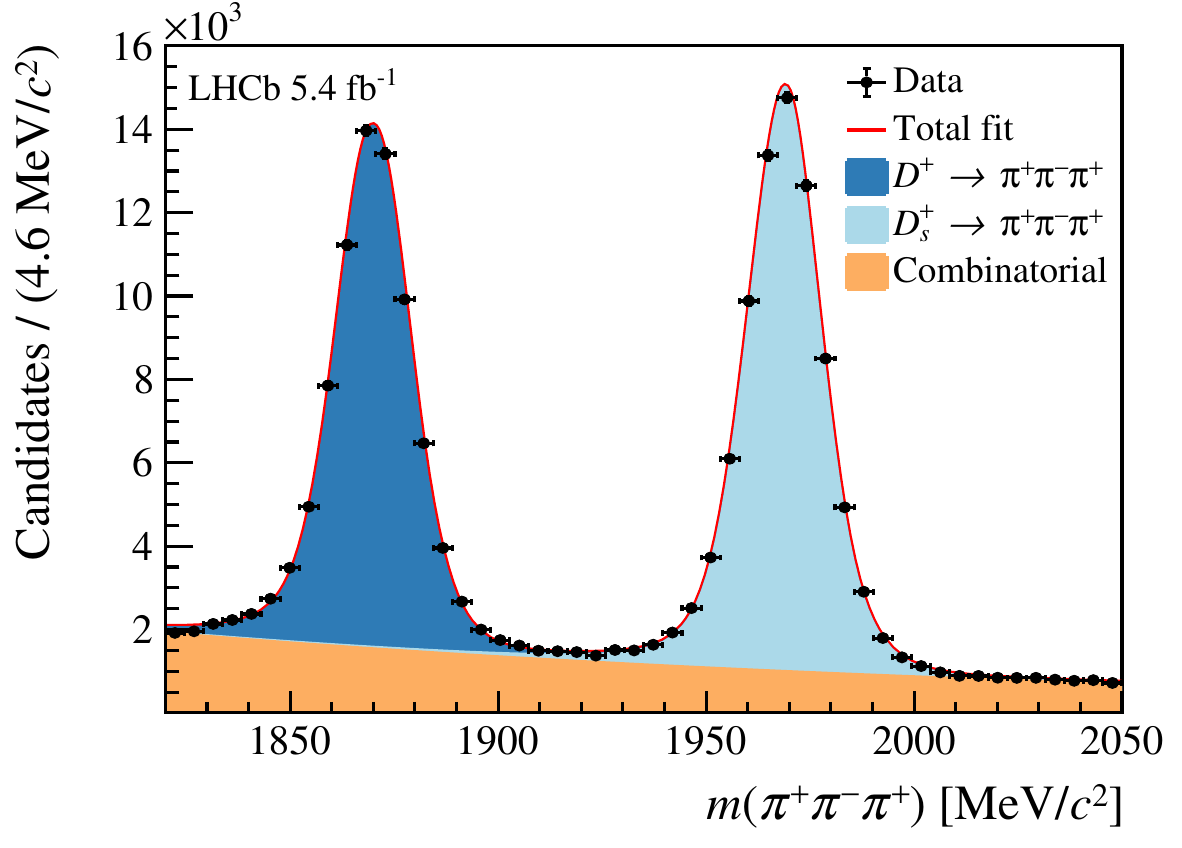}
\caption{Left: distribution of the $\pip\ell^{+}\ell^{-}$ invariant mass of candidates obtained from selecting final states with electrons of same charge (blue) or that fail the BDT requirement (orange). Fit results using a third-order polynomial are overlaid. The overall normalisation is chosen such that the two curves have the same area. Right: distribution of the invariant mass of $\pi^{+}\ell^{-}\ell^{+}$ candidates, after interpreting the leptons as pions, for candidates failing the electron PID criteria. The reversed PID sample contains predominantly \DPiPiPi events and a fraction of combinatorial events.}
\label{fig:bkgproxies}
\end{figure}

\begin{figure}[hb!]
\centering
\includegraphics[width=0.46\textwidth]{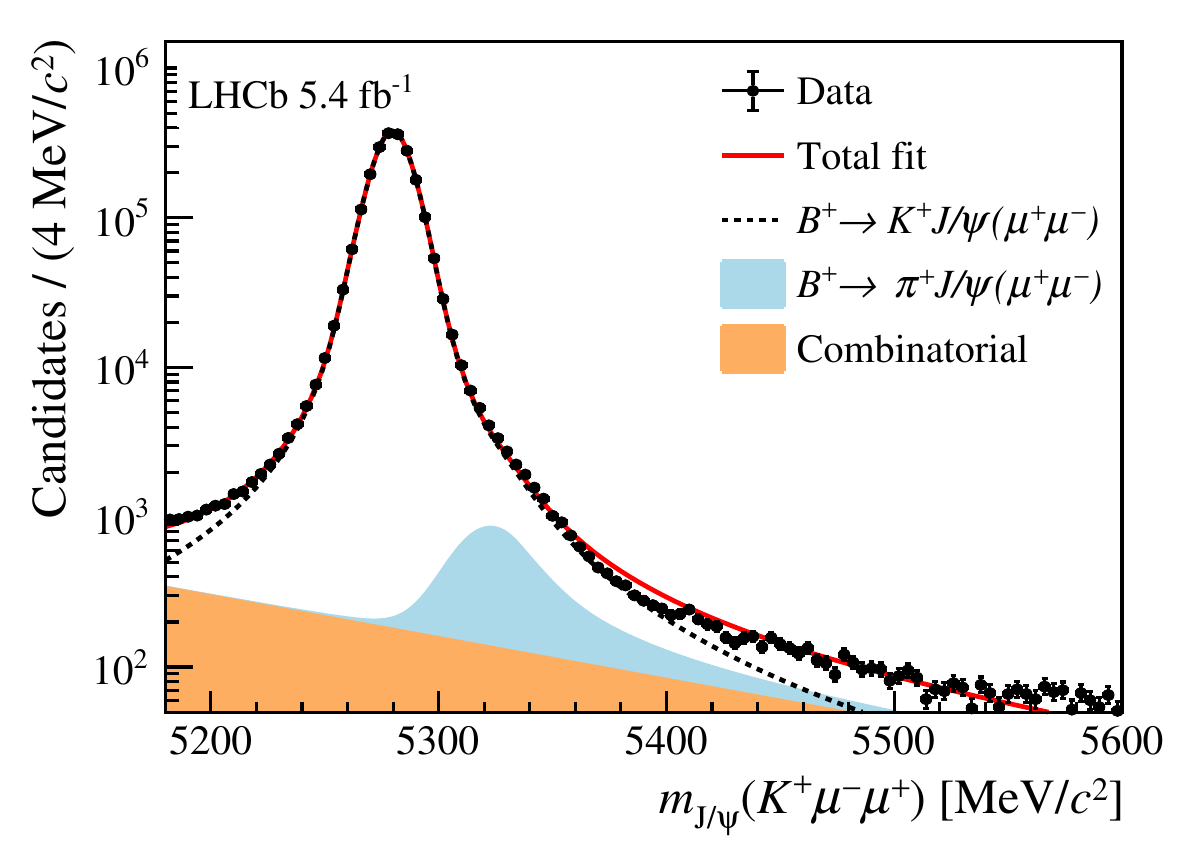}
\includegraphics[width=0.46\textwidth]{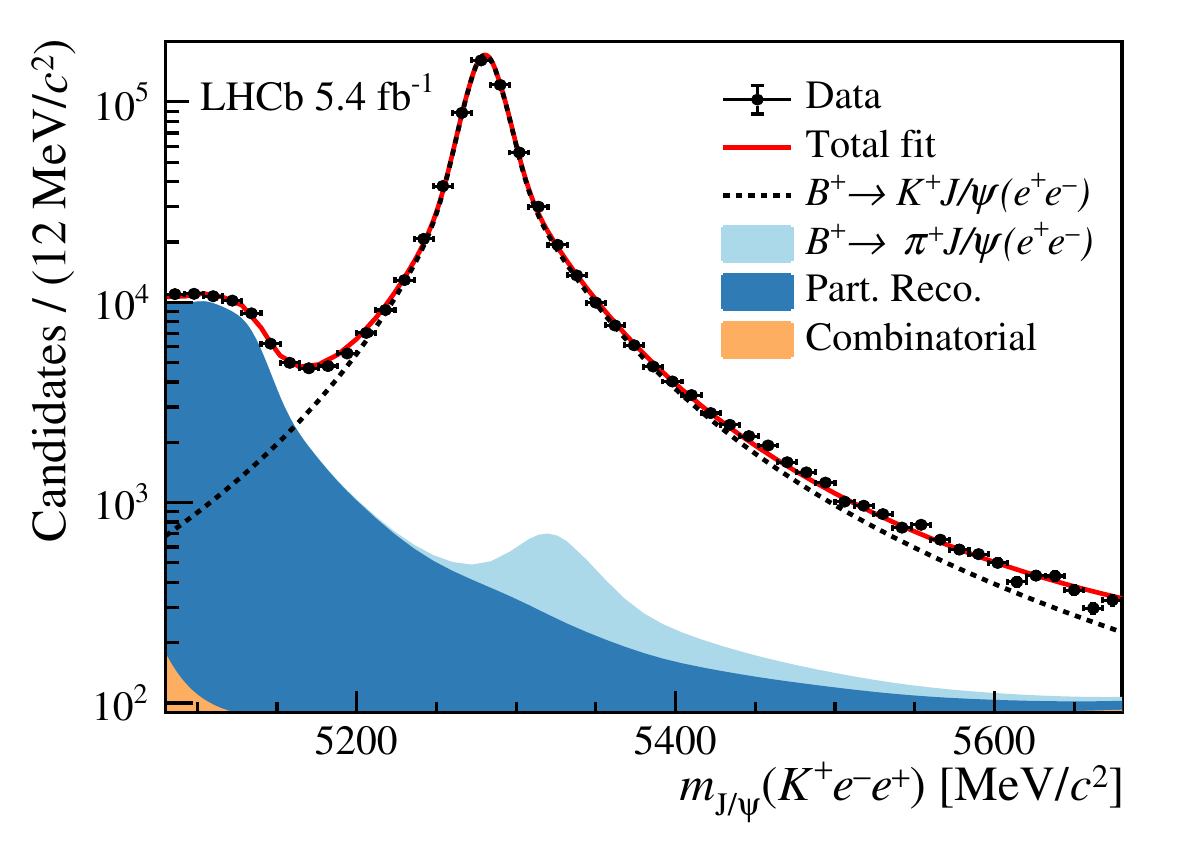}
\caption{Invariant mass distributions of the selected candidates for the (left) \BKJPsiee and (right) \BKJPsimumu modes, overlaid with the result of the fit. The dotted lines describe the signal contribution and the solid areas represent each of the background components described in the text.}

\label{fig:normfits}
\end{figure}

The combinatorial background is expected to follow an exponential shape as a function of the invariant mass in the case of the muon channel. However, due to the significantly worse resolution of the dilepton invariant mass in the electron channel, the $\phi$ mass constraint induces a strong variation of the efficiency as a function of \mPillconst near the boundaries of the $\Pill$ invariant mass selection requirement ($1770$\,--\,$2070\mevcc$) in the trigger. Therefore, the combinatorial background for the electron mode is parameterised by a third-order polynomial. This parameterisation is verified on a sample of candidates with leptons of same charge (same-sign) and a sample where the BDT classifier requirement is inverted to favour the combinatorial background, as shown in Fig.~\ref{fig:bkgproxies} (left).

The dominant peaking background originates from ${\DPiPiPi}$ decays, where two pions are misidentified as either two muons or two electrons. Due to the narrow $\phi$ mass window and the high rejection rate of the muon identification requirements, the contribution from this source in the muon mode is negligible. In the electron case, the size and shape of this background are obtained by reversing the electron identification requirements in data. A fit to the invariant-mass distribution of candidates that fail the PID criteria is used to distinguish \Dp and \Ds contributions from combinatorial background in the reversed electron ID sample. The invariant mass is computed using the relevant mass hypothesis for each track.
The result of this fit is shown in Fig.~\ref{fig:bkgproxies} (right). The \sPlot~\cite{Pivk:2004ty} weights resulting from this fit and the misidentification (misID) probabilities from data are employed to translate the sample lineshape and yield to the signal region, obtaining a model for the doubly misidentified background component. In the signal-yield determination fit, the yield of this background component is constrained using a Gaussian penalty function to the value obtained using the data-driven misID probabilities. The validity of this procedure is verified with a dedicated study described in Sect.~\ref{sec:systematics}.

The yields of the normalisation modes are determined via a maximum-likelihood fit to the \mKllconst distribution, as shown in Fig.~\ref{fig:normfits}. By analogy to the signal decays, the mass is obtained with the dilepton invariant mass constrained to the known \jpsi mass. This procedure vastly improves the resolution for both the electron and muon channels. The lineshapes for the normalisation channels are described with the same parameterisation as for the signal channels. Unlike the $\phi$ meson, the $\jpsi$ meson has a negligible natural width compared to the resolution, therefore the combinatorial background shape is unaffected by the \jpsi mass constraint, and is described with an exponential function. Due to the worse resolution of the electron channel, the dominant background in the electron mode comes from partially reconstructed \BuBdKpijpsi decays, where the pion is not included in the \Bp candidate. The lineshape of this component is obtained from simulation, while its yield is free to float in the fit. A small contribution from misidentified ${B^{+} \to \pi^{+} J/\psi(\to \ell^{+}\ell^{-})}$ background is included on the right-hand side of both muon and electron signal peaks. Their shape is obtained from simulation and their yield is constrained to the value obtained using the data-driven misID probabilities. The small disagreement near $5400\mevcc$ in the muon mode has a negligible impact on the signal-yield determination.

The total yields are obtained summing over the fit results for different trigger categories and run periods, and are summarised in Table~\ref{tab:yields}. In the muon channel, the correlation between the \DpPiPhi and \DsPiPhi yields is negligible, while in the electron channel the two yields are correlated by $30$\,--\,$45\%$, depending on the trigger category, due to the worse mass resolution for electrons.

\begin{table}[b]
\centering
\caption{Yields of the signal and normalisation decay modes obtained from the fits to the data. The quoted uncertainty accounts for both statistical and systematic effects.} \label{tab:yields}
\begin{tabular}{lc}
\toprule
Decay mode  &   Yield             \\
\midrule
\DpPiPhiee      &   $\phantom{0\,00}7\,460 \pm \phantom{0\,}140$ \\
\DpPiPhimumu      &   $\phantom{0\,0}43\,512 \pm \phantom{0\,}220$ \\
\DsPiPhiee      &   $\phantom{0\,0}16\,740 \pm \phantom{0\,}210$ \\
\DsPiPhimumu      &   $\phantom{0\,0}87\,022 \pm \phantom{0\,}300$ \\
\BKJPsiee  &   $\phantom{0\,}638\,600 \pm \phantom{0\,}900$ \\
\BKJPsimumu  &   $2\,187\,000 \pm 1\,500$      \\
\bottomrule
\end{tabular}
\end{table}

\section{Efficiency calculation}
\label{sec:efficiency}

The yields in Table~\ref{tab:yields} are used to compute \Rbothphipi according to the definition
\begin{equation}
\begin{split}
    \Rbothphipi & = \frac{N^{(d,s)}(\PiPhimumu)}{N^{(d,s)}(\PiPhiee)}\frac{\varepsilon^{(d,s)}(\PiPhiee)}{\varepsilon^{(d,s)}(\PiPhimumu)} \Big/r_{J/\psi}, \\ 
    % & = \beta_{e/\mu}\frac{N(H_{s}\mu\mu)}{N(H_{s}J/\psi(\mu\mu))}\frac{\varepsilon(H_{s}J/\psi(\mu\mu))}{\varepsilon(H_{s}\mu\mu)} \Big/\frac{N(H_{s}ee)}{N(H_{s}J/\psi(ee))}\frac{\varepsilon(H_{s}J/\psi(ee))}{\varepsilon(H_{s}ee)} \\
\end{split}
\end{equation}\label{eq:rphipi}

\noindent where $N^{(d,s)}(X)$ and $\varepsilon^{(d,s)}(X)$ are the yields and efficiencies to detect the decay $X$ initiated by a $D_{(s)}^{+}$ meson. The relative electron to muon efficiencies for signal and normalisation channels are determined from simulation. The efficiencies for each run period are obtained from a direct sum over the efficiency in each trigger category. The total efficiency for each channel is obtained by performing a luminosity-weighted average in the different run periods and are combined with the yields in Table~\ref{tab:yields} to derive the final result.

The simulation has several data-driven corrections applied, largely following the procedure used for the $R_K$ measurement described in Ref.~\cite{LHCb-PAPER-2021-004}, where each correction is applied with respect to the preceding one.

Firstly, the tracking efficiency for electrons is determined using a tag-and-probe method applied to a $\BKJPsiee$ sample~\cite{LHCb-DP-2019-003}. The difference in tracking efficiency between data and simulated samples, as a function of the pseudorapidity, the azimuthal angle $\phi$ and the \pt of the probe particle, is corrected for. The effect of these corrections is less than 1\%, reflecting the good agreement between simulation and data.

The PID efficiency is obtained from simulation using a set of control channels in data that can be efficiently selected without the application of PID criteria. These samples include the decays $\Dstarp\to \Dz \pip$ for hadron identification and $h\to \ell$ misID and ${\BKJPsimumu}$ for muon identification. The PID distribution in data is obtained using the \sPlot procedure~\cite{LHCb-DP-2018-001} as a function of the muon transverse momentum and pseudorapidity. The fraction of these distributions above the PID selection criteria are then used as PID efficiencies determined from data. For the electron PID response, the efficiencies are determined by refitting the signal ${\BKJPsiee}$ yield before and after the PID selection in bins of kinematic variables and bremsstrahlung recovery category, therefore taking into account differences in the background shape as a function of the PID response. The correction due to the assumption of factorisation of per-electron PID efficiencies is determined using simulation and is found to have a $<$0.1\% effect on the overall efficiency.

An initial set of corrections that are designed to account for the mis-modelling of the production kinematics and vertex quality in the simulation are applied. These corrections are obtained using the muon channels and applied to both muon and electron channels after the PID and tracking corrections have been applied. The corrections comprise a two-dimensional correction as a function of the \Bp and $D^{+}_{(s)}$ transverse momentum and pseudorapidity, followed by a one-dimensional correction to the quality of the fits to the \Bp and $D^{+}_{(s)}$ decay vertices. These corrections are applied iteratively to account for the correlation between the vertex quality and the production kinematics. Only two iterations are needed to obtain a satisfactory agreement in all three variables. The corrections to the efficiencies discussed below are obtained after applying these initial corrections to the simulation.

The largest efficiency difference between muons and electrons in the experiment originates from the trigger, where the \pt requirement for electrons is significantly more restrictive than for muons. Even when combining the three electron trigger categories, the total trigger efficiency of the electron channel is around 20\% of the muon channel. 
The normalisation channel is used to compare the trigger response between data and simulation, via a tag-and-probe method. For example, the kaon from the $\BKJPsiee$ decay, referred to as the tag, is required to pass a set of trigger requirements. The efficiency of the electron trigger is then determined by the fraction of candidates where either of the two electrons is also selected by the trigger. This efficiency is compared between data and simulation as a function of the maximum \pt of the two electrons, which is expected to be the kinematic variable on which the electron trigger efficiency most strongly depends. The efficiency for the hadron and muon triggers are determined in a similar way, as a function of the \pt of the kaon or muon. The discrepancy between measured and simulated efficiencies to trigger independently of the signal, is corrected differently, depending on the particle species that fired the trigger. For events triggered by any muon or hadron that are not part of the signal candidate, the trigger efficiency is parametrised in terms of the transverse momentum of the $B$ meson, while for events triggered by an electron or a photon which are not part of the signal, the maximum \pt of the dilepton pair is used.

A final set of corrections is obtained to correct for any residual mismodelling of the efficiencies as a function of the ${\Bp (D^{+}_{(s)})}$ production kinematic and geometrical quantities. Similar to the initial corrections, an iterative reweighting procedure is applied as a function of the ${\Bp (D^{+}_{(s)})}$ \pt and pseudorapidity, the ${\Bp (D^{+}_{(s)})}$ vertex-fit quality and the $\chisq_{\rm{IP}}$ with respect to the associated PV, where  $\chisq_{\rm{IP}}$ is defined as the difference in the vertex-fit \chisq of the PV reconstructed with and without the considered particle. These corrections are obtained using the muon channels, and are assumed to be independent of the lepton species and trigger category. This set of corrections is obtained after having accounted for the mismodelling of the PID and trigger efficiencies in the electron modes as well, enabling   evaluation of the systematic uncertainty due to these assumptions.

Finally, the dilepton invariant mass distribution is compared between data and simulation for ${\BKJPsiee}$ candidates. The dilepton invariant mass distribution of simulated events is fitted with a double-sided Crystal Ball~\cite{Skwarnicki:1986xj} function. This shape is then used to fit the data distribution with the mean and scale parameters, representing the mismatch between simulated and measured \qsq resolution, allowed to vary in the fit. These resolution parameters are determined as a function of the number of recovered bremsstrahlung photons and the trigger category. Overall, the resolution in the data is found to be $10$\,--\,$20\%$ worse than in simulation, with a shift in the peak position of around $2$\,--\,$10\mevcc$, depending on the trigger category. This resolution correction is applied to the \qsq and ${\Bp (D^{+}_{(s)})}$ invariant mass distributions of both normalisation and signal channels.  
This correction affects the shape of invariant mass distributions and therefore the efficiency of the invariant-mass requirements. The correction is derived assuming that the smearing parameters are independent of the electron pair momentum and the detector occupancy.
Due to the stringent requirement on the \Pill invariant mass in the trigger, assumptions associated with the resolution in the electron channel constitute one of the largest systematic uncertainties in the measurement.

\begin{figure}[tb]
\centering
\includegraphics[width=0.46\textwidth]{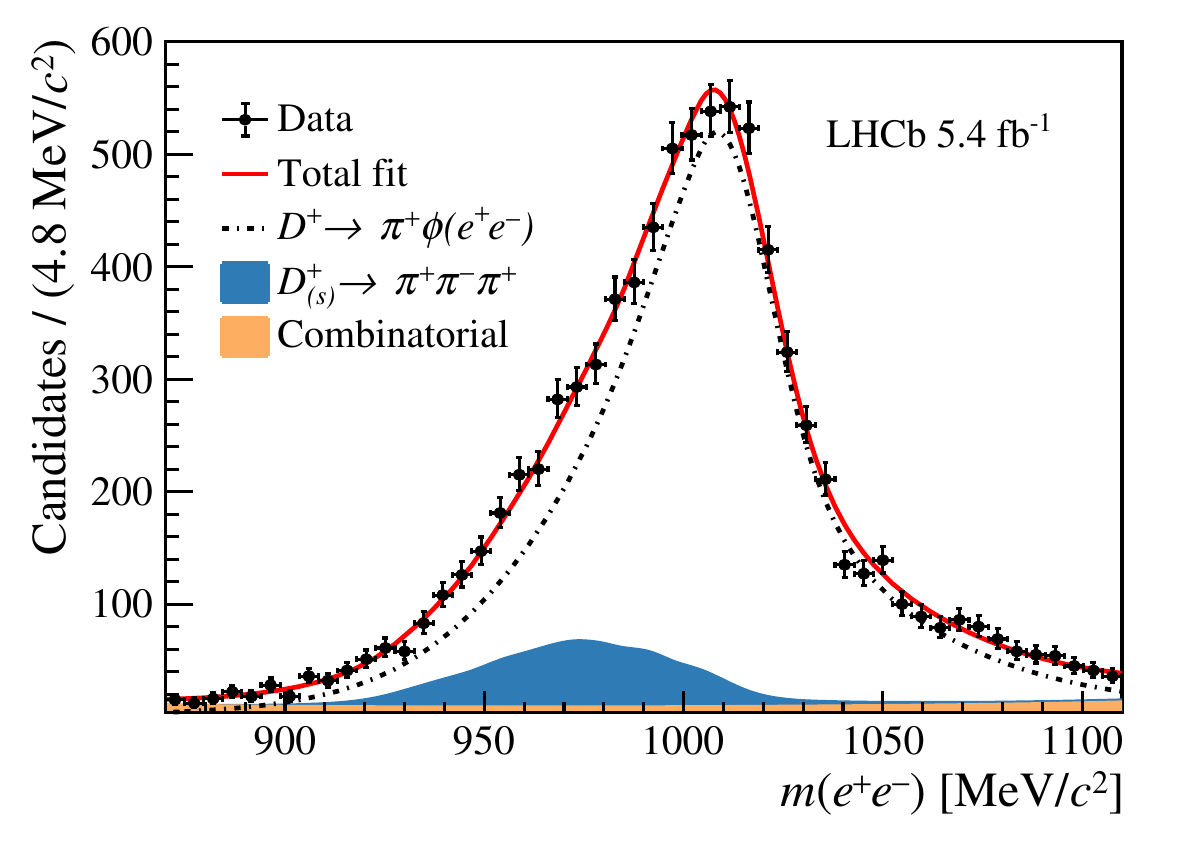}
\includegraphics[width=0.46\textwidth]{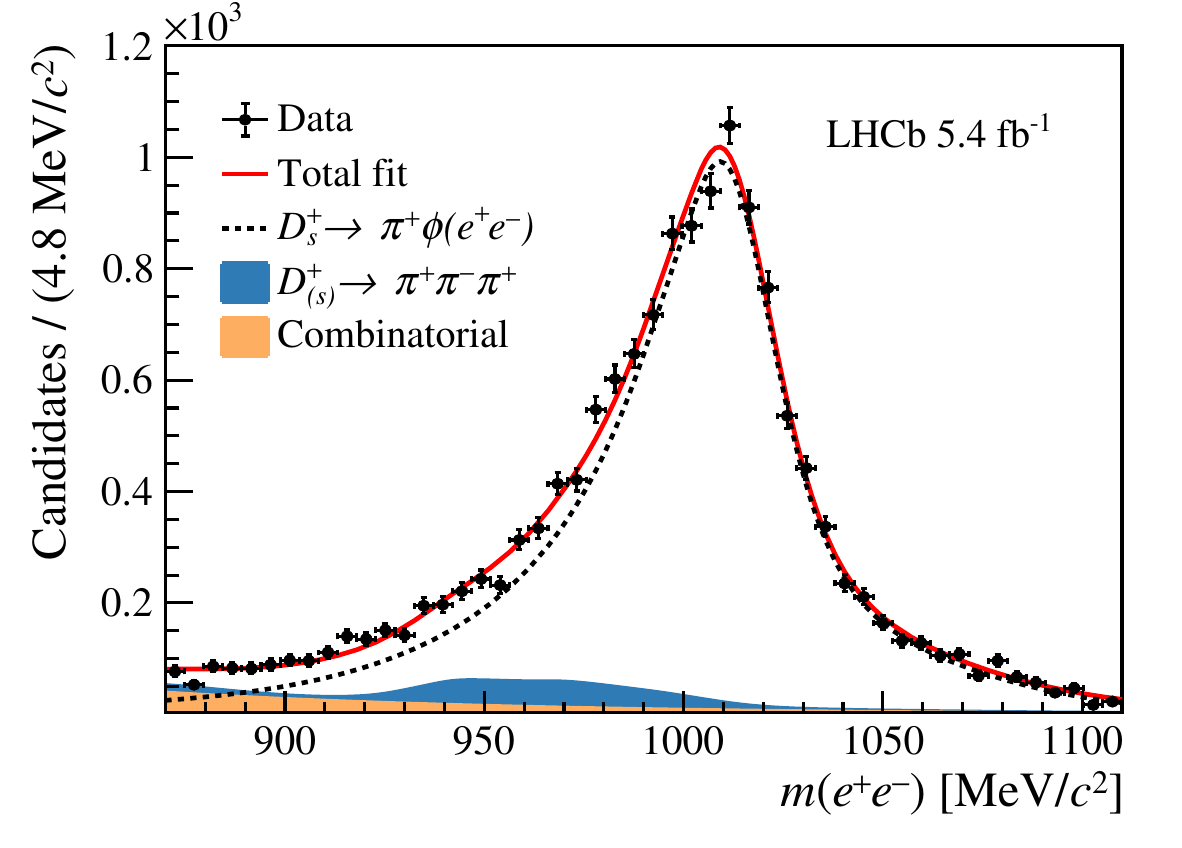}
\includegraphics[width=0.46\textwidth]{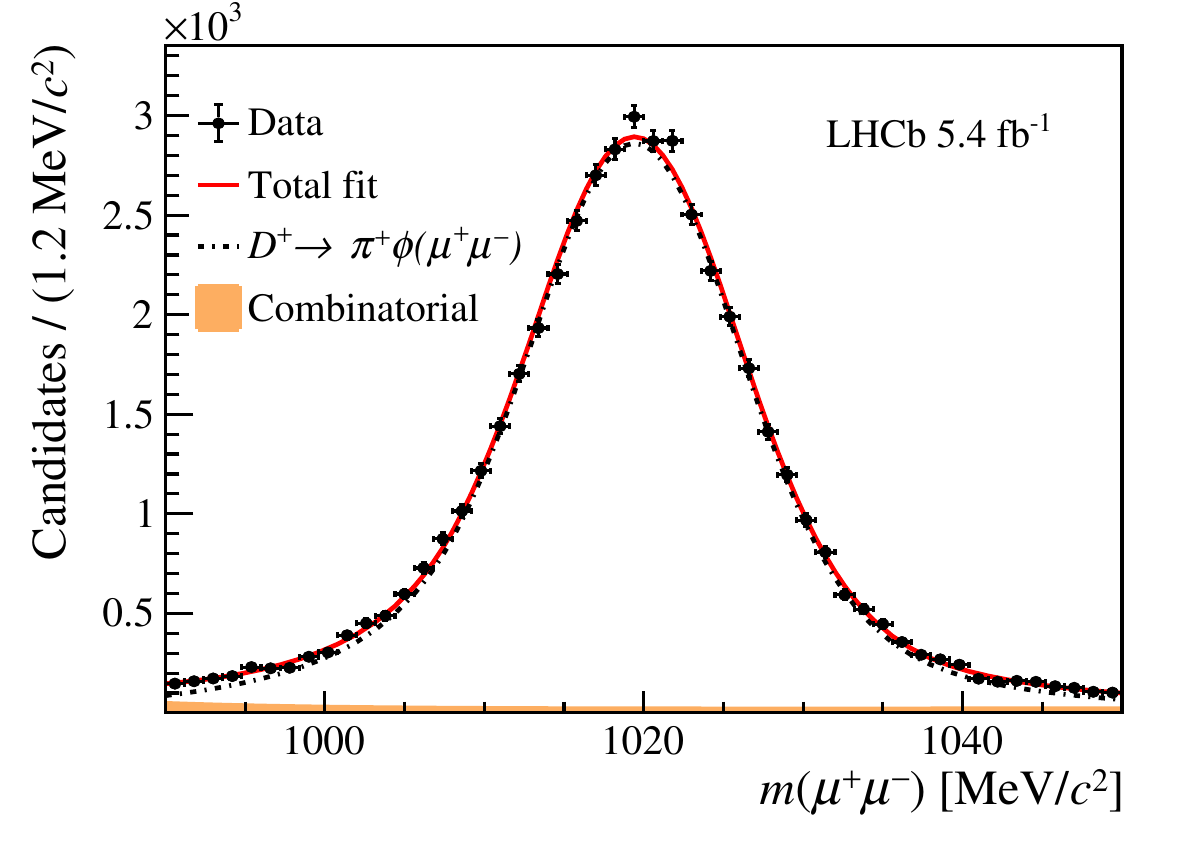}
\includegraphics[width=0.46\textwidth]{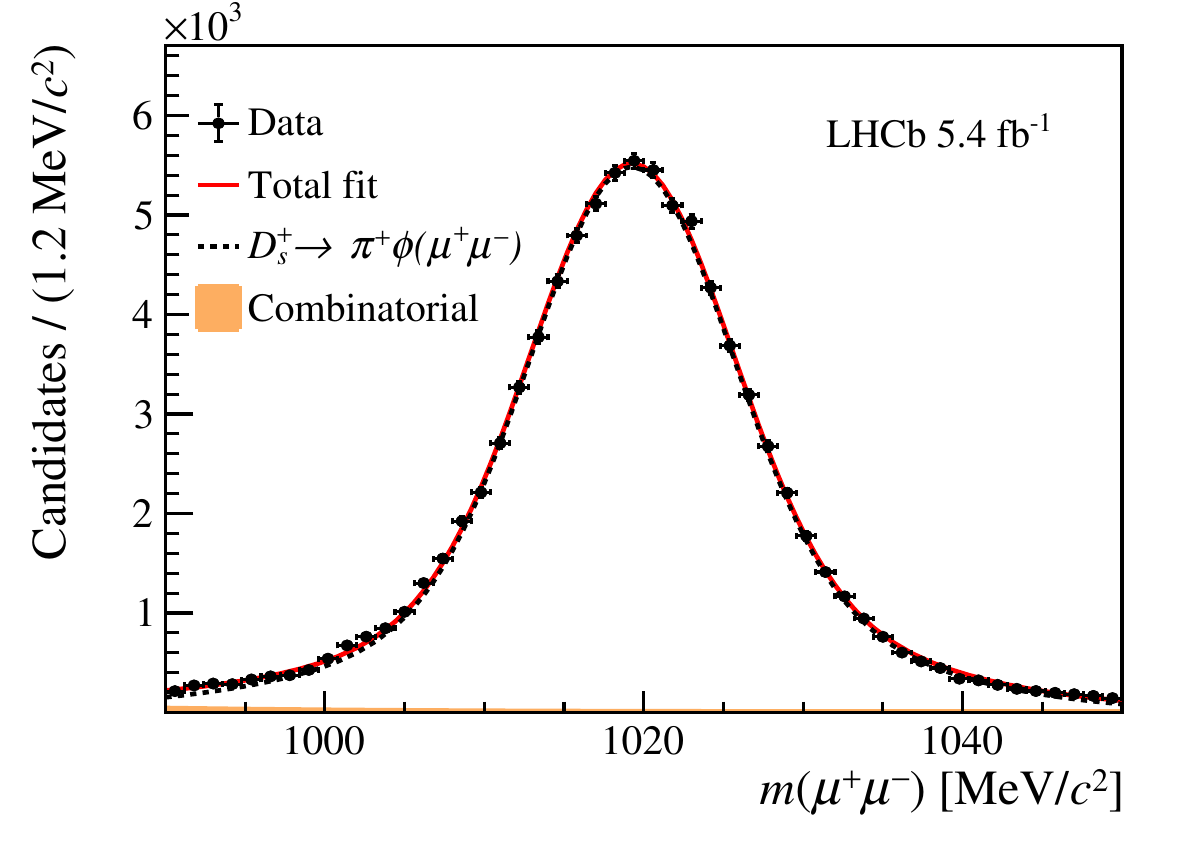}
\caption{Distributions of (top) $m(e^{+}e^{-})$ and (bottom) $m(\mu^{+}\mu^{-})$ for candidates selected in the signal region of the (left) \Dp (right) and \Ds $\phi$-constrained invariant mass. The dotted lines describe the signal contribution and the solid areas represent each of the background components described in the text.}
\label{fig:phimassfits}
\end{figure}

A fit to the dilepton mass distribution without any kinematic constraints is shown in Fig.~\ref{fig:phimassfits} for signal candidates selected in a narrow \mPillconst window around the known \Dp and \Ds masses. The signal shape is divided into categories based on the number of bremsstrahlung photons recovered, each modelled by a double-sided Crystal Ball function. The parameters of these distributions are obtained from a fit to the smeared distributions of simulated candidates. The agreement between the shapes obtained from simulation and the data provides a validation of the resolution description at low \qsq.

\section{Systematic uncertainties}
\label{sec:systematics}

A summary of the systematic uncertainties is given in Table~\ref{tab:syst}. Systematic uncertainties associated with the assumptions in the analysis are determined using variations of the result under alternative methods. The standard deviation of these variations is assigned as a systematic uncertainty. The systematic sources are assumed to be uncorrelated. 

\begin{table}[tb!]
  \centering
  \caption{Relative systematic uncertainties in percent affecting the measurements \Rdphipi and \Rsphipi. }
    \begin{tabular}{l c c}
      \toprule
      \noalign{\vskip 1mm}
      Source & \Rdphipi [\%] & \Rsphipi [\%] \\
      \midrule
      \noalign{\vskip 1mm}
Resolution on \qsq                              & 4.0 & 3.9 \\ 
Event multiplicity                              & 2.7 & 2.7 \\ 
Simulation reweighting                          & 1.5 & 1.2 \\
Combinatorial background shape parametrisation  & 1.5 & 1.0 \\
PID                                             & 0.8 & 0.8 \\ 
Finite size of control samples                  & 0.8 & 0.6 \\ 
Trigger                                         & 0.3 & 0.3 \\ 
Tracking                                        & 0.1 & 0.1 \\ 
Background from doubly misidentified electrons  & 1.1 & 0.1 \\ \midrule
      \noalign{\vskip 1mm}
Total                                           & 5.5 & 5.1 \\ 

      \bottomrule
    \end{tabular}
  \label{tab:syst}
  \end{table}

The assumptions used to correct the imperfect simulation of the dielectron \qsq resolution lead to the largest systematic uncertainty in the measurement.
The nominal resolution smearing parameters are assumed to be independent of the electron momenta. To account for the systematic uncertainty arising from this assumption, the simulated \qsq and ${\Bp (D^{+}_{(s)})}$ distributions are corrected by measuring the resolution in data for the normalisation channel as a function of the electron pair momentum. In addition, an alternative resolution smearing procedure is used, which depends on the number of hits reconstructed in the SPD. The sample is divided into three equally populated regions according to the number of SPD hits, and separate smearing parameters are obtained for each region. Additional uncertainties arising from the limited precision of the resolution parameters are found to have a negligible effect.

The particle multiplicity is mismodelled in the simulation. For example, there are around 10\% fewer tracks per event and around 30\% fewer hits in the SPD reconstructed in simulation compared to the data. A systematic correction associated to this mismodelling is determined by adding to the initial set of corrections a separate reweighting of three different multiplicity proxies: the number of fully reconstructed tracks, the number of primary vertices reconstructed in the VELO and the number of hits reconstructed in the SPD. Although only weakly correlated between different lepton flavours, this leads to a large systematic uncertainty due to the mismodelling of the SPD multiplicity and the different production mechanisms for the signal and normalisation channels.

The final set of corrections is obtained from the muon channels and is applied coherently to both electron and muon channels. A systematic uncertainty based on the assumption that the corrections are independent of lepton flavour is determined by obtaining the corrections from the \BKJPsiee channel instead. An additional set of mixed corrections is also derived, whereby the production corrections originate from the muon control channels and the vertex quality and IP \chisq originate from the electron control channels, and are then applied to the signal channel. The impact of the trigger on the weights applied to correct the simulation is determined by varying the trigger requirement for the samples used in those corrections.  

Three alternative methods for extracting the signal yields are adopted to determine the systematic uncertainty arising from the combinatorial background parameterisation for the electron mode. Firstly, the coefficients describing the combinatorial shape in the fit are constrained to the values obtained from two independent proxies: a sample where the two reconstructed leptons have the same charge and a sample where the BDT requirement is reversed. Secondly, the assumption arising from the truncation of the polynomial is relaxed by increasing it to fourth order. 

Three systematic sources are considered for the determination of the PID efficiencies: the binning scheme of the PID response as a function of the kinematic variables; the dependence of the PID response on the trigger; and the assumption that the PID response factorises between the two electrons. As the PID selection is highly efficient, these assumptions result in a small systematic uncertainty. A comparably small uncertainty is associated with the tracking efficiency corrections.

The limited size of the control samples used to derive corrections to the simulated efficiencies gives rise to an associated systematic uncertainty. The impact of this is determined by bootstrapping~\cite{10.1214/aos/1176344552} the relevant samples and re-computing the associated efficiencies. The 68\% envelope of these repeated efficiency determinations is assigned as the systematic uncertainty.

The systematic uncertainty associated with the trigger corrections originates from the dependence on the requirement on the tag lepton used in the tag-and-probe method. This is determined by varying this tag requirement, where the largest effects are seen in the hadronic trigger category. The assumption that the trigger efficiencies factorise between the two electrons is also determined by parameterising the trigger response as a function of the distance between the electron clusters in the ECAL.

\begin{figure}[t]
\centering
\includegraphics[width=0.48\textwidth]{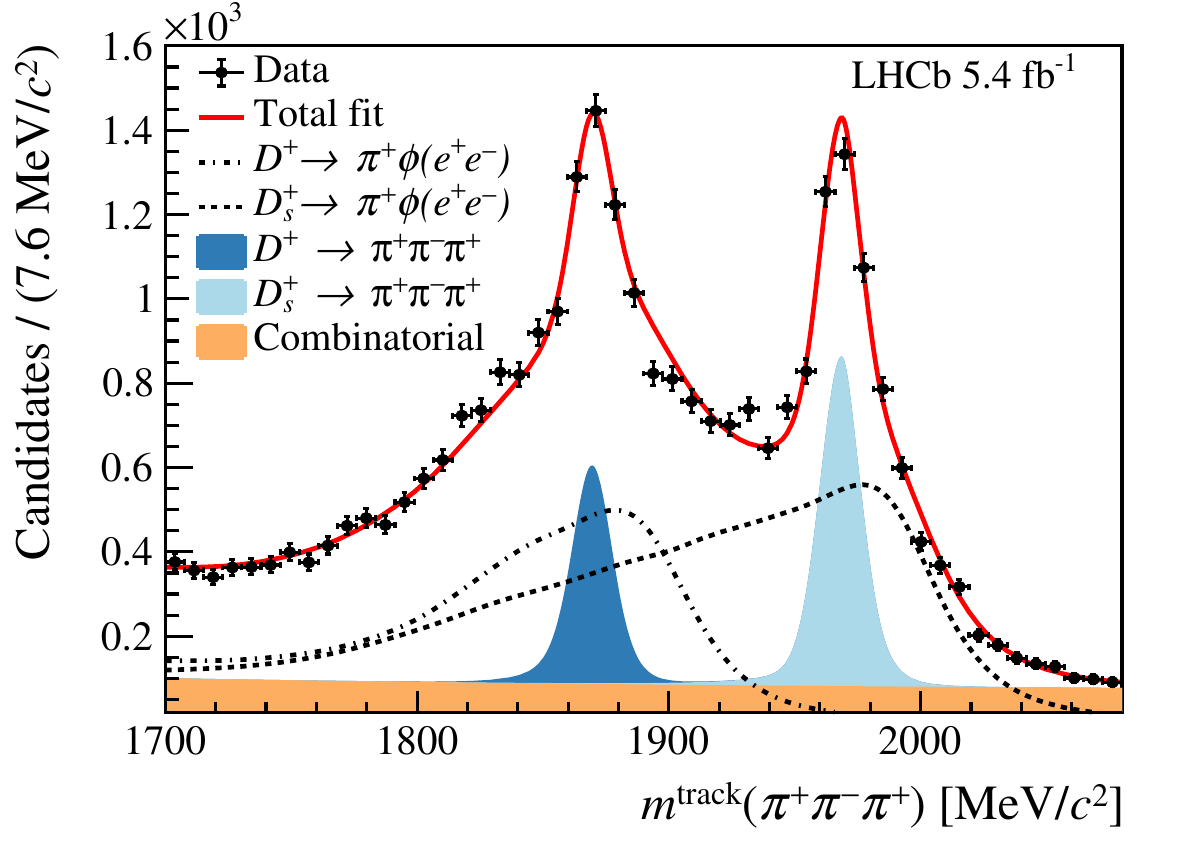}
\caption{Distribution of the invariant mass computed 
under the pion mass hypothesis for the dielectron pair using track information only, for candidates passing the electron PID criteria. The fit projection is superimposed, with dashed and dotted-dashed lines describing the signal contributions, while solid light and dark blue areas represent misidentified $\mbox{\DPiPiPi}$ components. In this variable the latter are clearly visible in data passing the electron PID criteria.}
\label{fig:mpipipi_normpid}
\end{figure}

In the nominal signal-yield determination, the estimate of the doubly misidentified background relies on the translation of this background source from the reversed PID region to the signal PID region using the known PID efficiencies. An alternative procedure that does not rely on this method is used to extract the signal yields, and the difference between these results is taken as a systematic uncertainty. In the alternative signal extraction procedure, the expected \DPiPiPi contamination is obtained from a fit to the $\pi\pi\pi$ invariant-mass distribution for signal candidates selected in the signal PID region, such that no translation across PID regions is needed. The result of the fit is shown in Fig.~\ref{fig:mpipipi_normpid}. In this alternative approach, the contamination of doubly misidentified background events is constrained to the one extracted from the nominal PID region fit. This leads to a comparitavely larger systematic uncertainty for the \DpPiPhi mode compared to the \DsPiPhi decay as the misID background peaks in the vicinity of the \Dp mass. The small size of this systematic uncertainty confirms the accuracy of data-driven estimates of the misidentified $h \to e$ background sources in LFU analyses at low \qsq.

\section{Results and conclusion}
\label{sec:results}

The final results are
  \begin{align*}
  \RdphipiResult, \\
  \RsphipiResult.  
  \end{align*}
A combination of the ratios from the two charm hadron decays is obtained by performing a weighted average of the two results, taking into account the correlations
  \begin{displaymath}
  \RphipiResult.
  \end{displaymath}
This result is combined with the existing $\phi\to e^+ e^-$ branching fraction measurement of $(2.979 \pm 0.033) \times 10^{-4}$~\cite{PDG2022} to obtain a measurement of 
\begin{displaymath}
{\cal B} (\phi\to \mu^+ \mu^-) = (3.045 \pm 0.049 \,({\rm stat}) \pm 0.148 \,({\rm syst})) \times 10^{-4},
  \end{displaymath}
This is the most precise measurement of this branching fraction to date.

\begin{figure}[htb]
\centering
\includegraphics[width=0.46\textwidth]{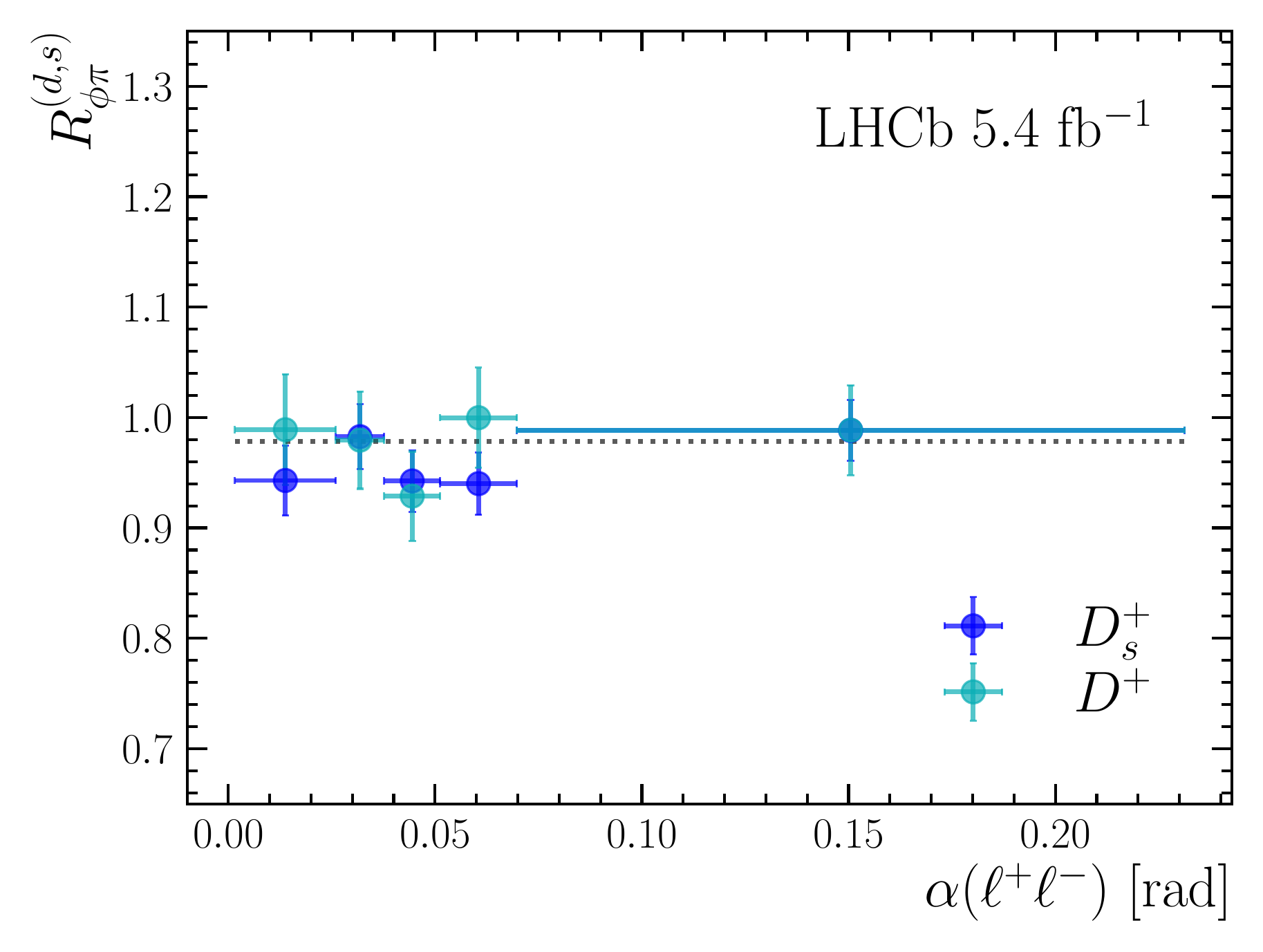}
\includegraphics[width=0.46\textwidth]{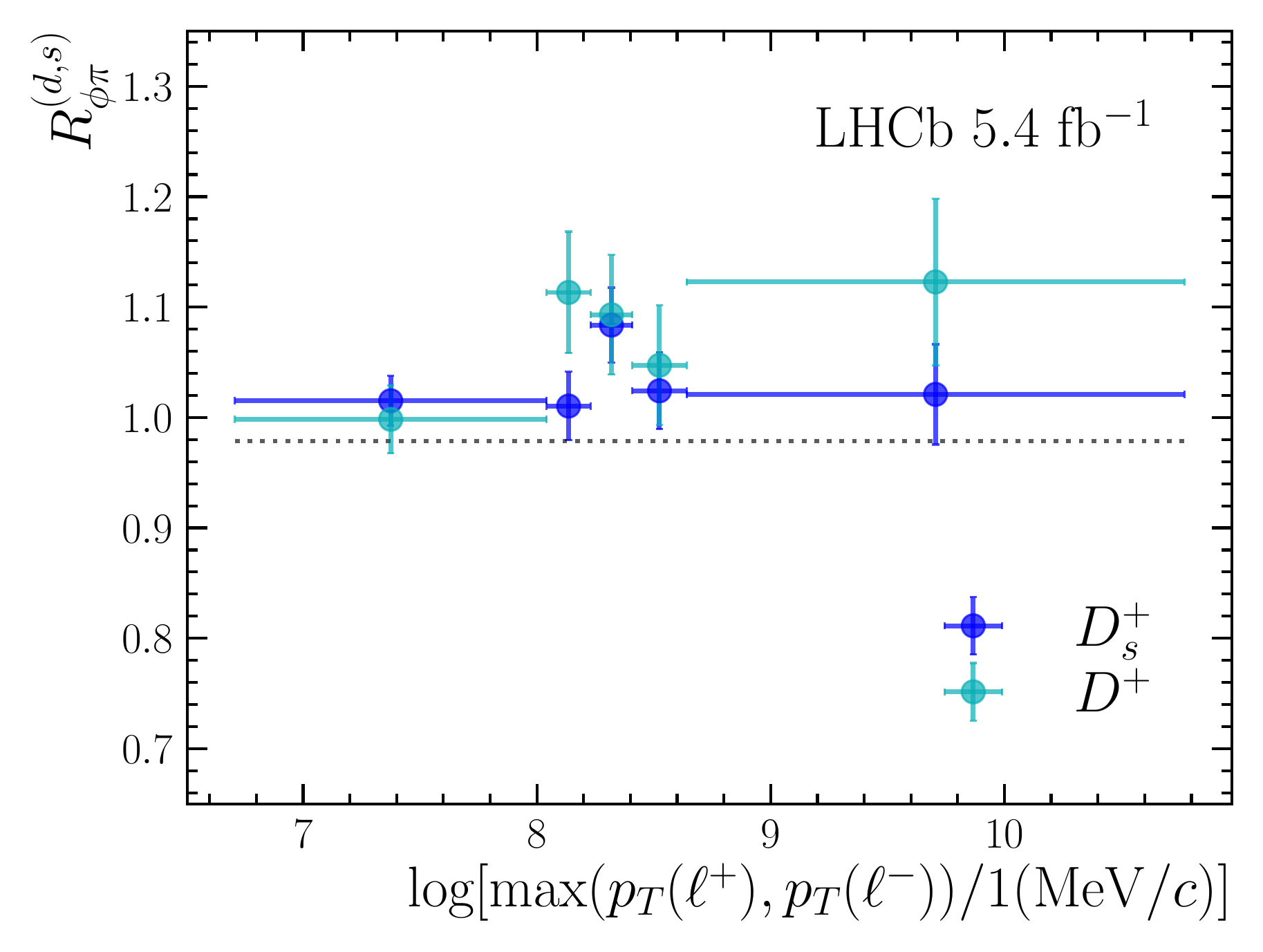}
\includegraphics[width=0.46\textwidth]{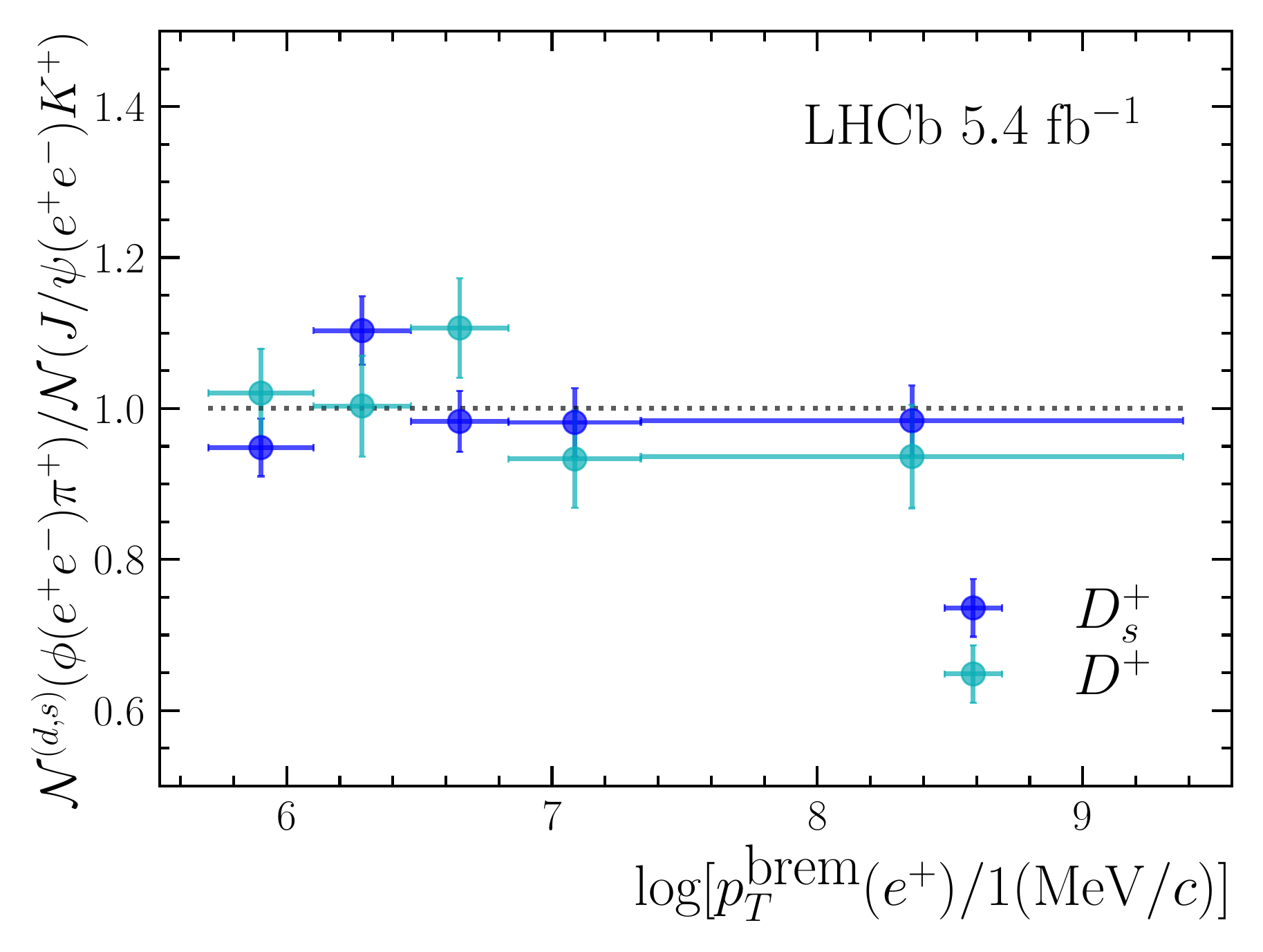}
\includegraphics[width=0.46\textwidth]{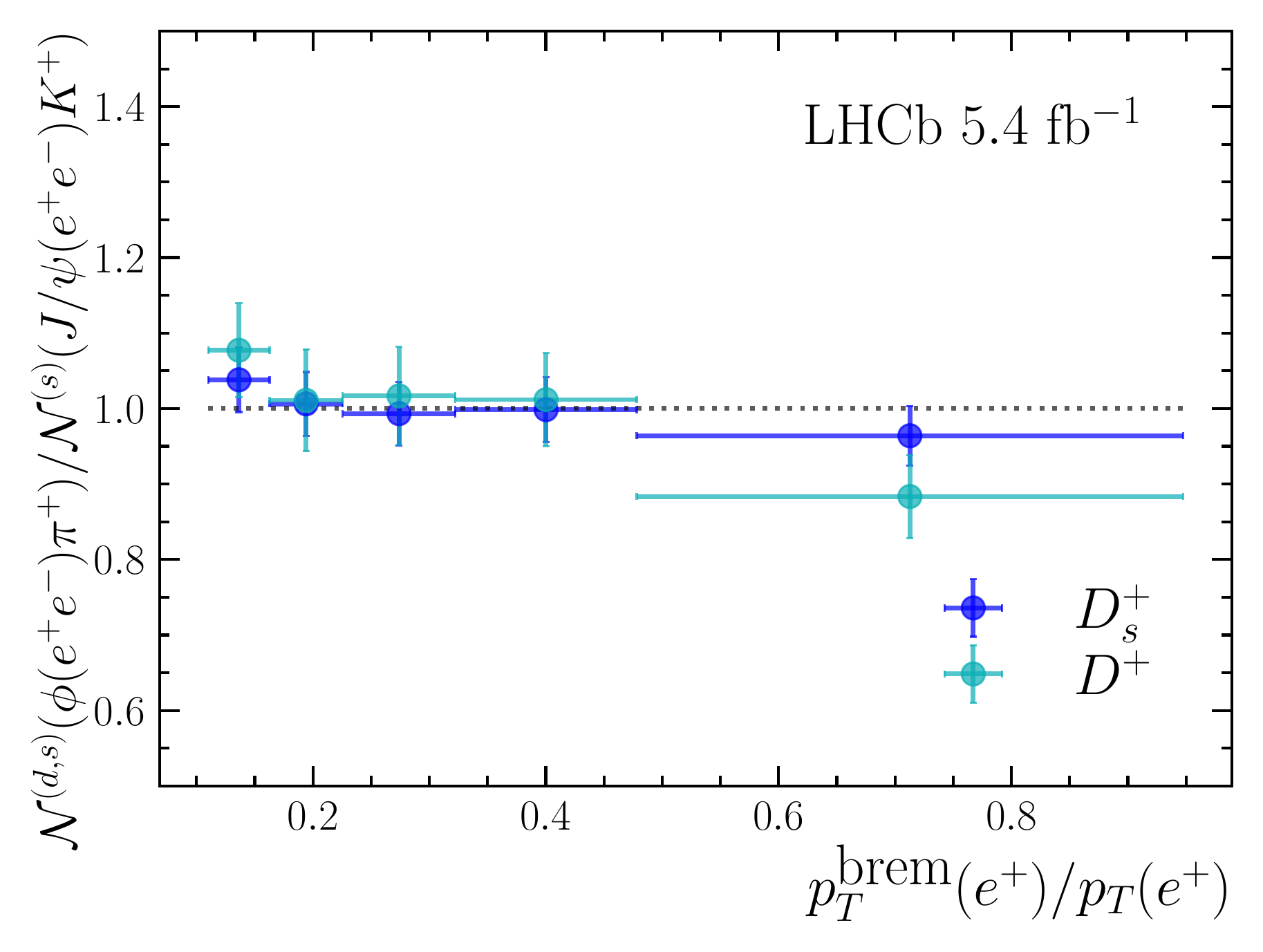}
\caption{Measured \Rbothphipi as a function of the (top left) angle between the leptons $\alpha(\ell^+, \ell^-)$, and (top right) the maximum \pt of the leptons. The structure observed at intermediate maximum \pt is related to the trigger threshold. The errors shown are statistical only. Also shown is the ratio of efficiency-corrected ${\DPiPhi}$ and ${\BKJPsiee}$ yields as a function of the (bottom left) transverse momentum recovered with the bremsstrahlung recovery algorithm and (bottom right) its fraction with respect to the total transverse momentum of the electron. The flatness of these distributions indicates that the bremsstrahlung recovery algorithm is well reproduced in simulation at low \qsq. The normalisation of these distributions is arbitrary and the uncertainties displayed are statistical only.
}
\label{fig:Rphiflatness1}
\end{figure}

These results are compatible with previous measurements \cite{PDG2022} and with lepton flavour universality at less than one standard deviation. Systematic uncertainties related to the efficiency determination are larger than those in the $R_K$ and $R_{K^{*}}$ analyses~\cite{LHCb-PAPER-2022-045,LHCb-PAPER-2022-046} due to the different production mechanism between the signal and normalisation channels and restrictive invariant mass requirements in the trigger. This restriction only affects the charm decays analysed in this measurement and can be improved in the future. In addition, the variation of the measured value as a function of several kinematic variables is shown in Fig.~\ref{fig:Rphiflatness1}. No significant dependence on these variables is observed. Assuming the fluctuations seen in these variables are attributed to a systematic effect, the maximum deviation in the efficiency is found to be 2.5\%, which shows that the efficiency of electrons can be controlled at least at this level for low-mass di-electron signatures.

Similar to the $R_{K^{(*)}}$ measurements~\cite{LHCb-PAPER-2022-045,LHCb-PAPER-2022-046}, the single ratio of the normalisation channel $\rjpsi\xspace$ is determined to cross-check the efficiency correction procedure. This is found to be compatible with the SM at the level of $0.6$ sigma and with previous analyses. Such agreement is expected, as a very similar correction procedure is applied in this case. 

In summary, the branching fraction ratio $\mathcal{B}$(\phimumu)/$\mathcal{B}$(\phiee) is determined with $\DpPiPhi$ and $\DsPiPhi$ decays, using a dataset corresponding to an integrated luminosity of 5.4\invfb of $pp$ collision data collected with the LHCb experiment. The results are consistent with the Standard Model and with previous measurements. This is the first lepton flavour universality test in $\phi$ meson decays at the LHCb experiment. It probes a unique kinematic region and is crucial for understanding the experimental features of low-mass dileptons in the LHCb environment.

% Do not include this in any draft (just for information in the template)
%\input{acknowledgements_intro}
% Comment this in for paper drafts; do not include this in analysis note, conference and figure reports
\section*{Acknowledgements}
%
% These Acknowledgements valid from 3-May-2019
%
\noindent We express our gratitude to our colleagues in the CERN
accelerator departments for the excellent performance of the LHC. We
thank the technical and administrative staff at the LHCb
institutes.
We acknowledge support from CERN and from the national agencies:
CAPES, CNPq, FAPERJ and FINEP (Brazil); 
MOST and NSFC (China); 
CNRS/IN2P3 (France); 
BMBF, DFG and MPG (Germany); 
INFN (Italy); 
NWO (Netherlands); 
MNiSW and NCN (Poland); 
MCID/IFA (Romania); 
%MSHE (Russia); 
MICINN (Spain); 
SNSF and SER (Switzerland); 
NASU (Ukraine); 
STFC (United Kingdom); 
DOE NP and NSF (USA).
We acknowledge the computing resources that are provided by CERN, IN2P3
(France), KIT and DESY (Germany), INFN (Italy), SURF (Netherlands),
PIC (Spain), GridPP (United Kingdom), 
%RRCKI and Yandex LLC (Russia), 
CSCS (Switzerland), IFIN-HH (Romania), CBPF (Brazil),
and Polish WLCG (Poland).
We are indebted to the communities behind the multiple open-source
software packages on which we depend.
Individual groups or members have received support from
ARC and ARDC (Australia);
Key Research Program of Frontier Sciences of CAS, CAS PIFI, CAS CCEPP, 
Fundamental Research Funds for the Central Universities, 
and Sci. \& Tech. Program of Guangzhou (China);
Minciencias (Colombia);
EPLANET, Marie Sk\l{}odowska-Curie Actions, ERC and NextGenerationEU (European Union);
A*MIDEX, ANR, IPhU and Labex P2IO, and R\'{e}gion Auvergne-Rh\^{o}ne-Alpes (France);
%RFBR, RSF and Yandex LLC (Russia);
AvH Foundation (Germany);
ICSC (Italy); 
GVA, XuntaGal, GENCAT, Inditex, InTalent and Prog.~Atracci\'on Talento, CM (Spain);
SRC (Sweden);
the Leverhulme Trust, the Royal Society
 and UKRI (United Kingdom).

%\input{supplementary}

%\input{appendix}

% This should be taken out in the final paper
%\input{supplementary}

\addcontentsline{toc}{section}{References}
%\setboolean{inbibliography}{true}
\bibliographystyle{LHCb}
\bibliography{main,standard,LHCb-PAPER,LHCb-CONF,LHCb-DP,LHCb-TDR}

\newpage
% LHCb collaboration author list
% Data extracted on January 22nd, 2024 at 6:22pm for paper reference LHCb-PAPER-2023-038
\centerline
{\large\bf LHCb collaboration}
\begin
{flushleft}
\small
R.~Aaij$^{35}$\lhcborcid{0000-0003-0533-1952},
A.S.W.~Abdelmotteleb$^{54}$\lhcborcid{0000-0001-7905-0542},
C.~Abellan~Beteta$^{48}$,
F.~Abudin{\'e}n$^{54}$\lhcborcid{0000-0002-6737-3528},
T.~Ackernley$^{58}$\lhcborcid{0000-0002-5951-3498},
B.~Adeva$^{44}$\lhcborcid{0000-0001-9756-3712},
M.~Adinolfi$^{52}$\lhcborcid{0000-0002-1326-1264},
P.~Adlarson$^{78}$\lhcborcid{0000-0001-6280-3851},
C.~Agapopoulou$^{46}$\lhcborcid{0000-0002-2368-0147},
C.A.~Aidala$^{79}$\lhcborcid{0000-0001-9540-4988}
% ...

{\footnotesize \it

% ...
$^{35}$Nikhef National Institute for Subatomic Physics, Amsterdam, Netherlands\\
% ...
$^{44}$Instituto Galego de F{\'\i}sica de Altas Enerx{\'\i}as (IGFAE), Universidade de Santiago de Compostela, Santiago de Compostela, Spain\\
% ...
$^{46}$European Organization for Nuclear Research (CERN), Geneva, Switzerland\\
% ...
$^{48}$Physik-Institut, Universit{\"a}t Z{\"u}rich, Z{\"u}rich, Switzerland\\
% ...
$^{52}$H.H. Wills Physics Laboratory, University of Bristol, Bristol, United Kingdom\\
% ...
$^{54}$Department of Physics, University of Warwick, Coventry, United Kingdom\\
% ...
$^{57}$School of Physics and Astronomy, University of Glasgow, Glasgow, United Kingdom\\
$^{58}$Oliver Lodge Laboratory, University of Liverpool, Liverpool, United Kingdom\\
% ...
$^{66}$Syracuse University, Syracuse, NY, United States\\
% ...
$^{78}$Department of Physics and Astronomy, Uppsala University, Uppsala, Sweden, associated to $^{57}$\\
$^{79}$University of Michigan, Ann Arbor, MI, United States, associated to $^{66}$\\
% ...
}
\end{flushleft}

\end{document}